\documentclass{aa}  

\usepackage{graphicx}
\usepackage{txfonts}
\usepackage{hyperref}
\hypersetup{colorlinks=true,linkcolor=blue,citecolor=blue,filecolor=blue,urlcolor=blue}
\makeatletter
\renewcommand*\aa@pageof{, page \thepage{} of \pageref*{LastPage}}
\makeatother

\usepackage{natbib}
\usepackage{soul}
\usepackage{makecell}
\usepackage{lscape}                    
\usepackage{longtable}               
\usepackage{multirow}                
\usepackage{multicol}                
\usepackage{footnote}                
\usepackage{rotating}
\usepackage{threeparttable}
\usepackage{tabularx}
\usepackage{threeparttablex}
\usepackage{footmisc}
\usepackage{siunitx}

\newcommand{\CAT}{\textit{UltraCompCAT}}

\newcommand{\Fx}{$F_\mathrm{X}$}
\newcommand{\lx}{$L_\mathrm{X}$}

\newcommand{\Nh}{$N_{\rm H}$}

\newcommand{\Po}{$P_\mathrm{orb}$}

\newcommand{\Msun}{\mathrm{M}_{\odot}}
\newcommand{\lum}{\mathrm{erg~s}^{-1}}
\newcommand{\flux}{\mathrm{erg~cm}^{-2}~\mathrm{s}^{-1}}

\newcommand{\nh}{$\mathrm{cm}^{-2}$}

\newcommand{\chan}{\textit{Chandra}}

\newcommand{\xmm}{\textit{XMM-Newton}}

\newcommand{\rxte}{\textit{RXTE}}

\newcommand{\rosat}{\textit{ROSAT}}

\newcommand{\nicer}{\textit{NICER}}
\newcommand{\gaia}{\textit{Gaia}}

\begin{document}

   \title{UltraCompCAT: a comprehensive Catalogue of Ultra-Compact and Short Orbital Period X-ray Binaries}


   \author{M.~Armas~Padilla
          \inst{1,}\inst{2}
          \and
          J.~M.~Corral-Santana
          \inst{3}
          \and
          A.~Borghese
          \inst{1,}\inst{2}
          \and
          V.~A.~Cúneo
          \inst{1,}\inst{2}
          \and
          T.~Muñoz-Darias
          \inst{1,}\inst{2}
          \and
          J.~Casares
          \inst{1,}\inst{2}
          \and
          M.~A.~P.~Torres
          \inst{1,}\inst{2}
          }

   \institute{Instituto de Astrof\'isica de Canarias (IAC), V\'ia L\'actea s/n, La Laguna 38205, S/C de Tenerife, Spain\\
              \email{m.armaspadilla@iac.es}
         \and
             Departamento de Astrof\'isica, Universidad de La Laguna, La Laguna, E-38205, S/C de Tenerife, Spain
        \and
           European Southern Observatory, Alonso de C\'ordova 3107, Vitacura, Casilla 19001, Santiago de Chile, Chile 
             }

   \date{Received September 15, 1996; accepted March 16, 1997}

\abstract %
  {
  Ultracompact X-ray binaries (UCXBs) are a distinctive but elusive family of low-mass X-ray binaries (LMXBs) characterised by their tight orbits and degenerate donor stars. Here we present \CAT, the first online and comprehensive catalogue of UCXBs. The initial version of \CAT\ comprises 49 sources, including 20 `confirmed' UCXBs (those with a measured orbital period shorter than 80 min) and 25 systems that we label as `candidate' based on their multi-wavelength phenomenology. For completeness, we also include four LMXBs with orbital periods in the range of 80 to 120 min, since they might be related (e.g. close progenitors) or even part of the UCXB population that evolved towards longer periods. We discuss the orbital period and Galactic distribution of the catalogue's sample. We provide evidence for the presence of at least two separate groups of UCXBs. One formed by persistent systems with orbital periods shorter than 30 min and a second group of transient objects (70 \%) with periods in the range of 40 to 60 min. We show that the former group is dominated by sources formed in globular clusters, while the latter accounts for the (known) UCXB population in the Galactic field. We discuss the possible evolutionary channels for both groups.  
  
  }

    \titlerunning{UltraCompCAT}
   \maketitle
%

\section{Introduction}
\label{intro}

Low-mass X-ray binaries (LMXBs) are stellar systems composed of a black hole or a neutron star accreting material from a Roche lobe filling low-mass ($\lesssim 1~\Msun$) companion, the donor star. A few hundreds of these systems have been already discovered in our galaxy (e.g. \citealt{Liu2007}), including both persistently active X-ray sources and transient objects (e.g. \citealt{Casares2017}). However, thousands more are expected to exist, mainly as transient objects with recurrence periods longer than $\sim$ 50 years, the time scale sampled by X-ray astronomy (e.g. \citealt{Corral-Santana2016, Tetarenko2016,Bahramian2022}).

The orbital period (\Po) is arguably the most important parameter of a binary system. Among other things, it is a proxy for its size, and thus the physical properties of its components. Regardless of their persistent or transient nature, `classical' LMXBs can be defined as those with hydrogen burning donors (e.g. main sequence stars). They have \Po\ in the range of 80~min to tens of days. Contrastingly, systems with
\Po\ shorter than $\sim$ 60--80 min are thought to harbour hydrogen-poor companions, which do not burn hydrogen in their cores and may contain electron-degenerate matter \citep{Paczynski1981,Rappaport1982}. LMXBs harbouring this kind of donors are known as ultra-compact X-ray binaries (UCXBs; see also \citealt{Solheim2010}, for similar compact binaries with WD accretors). By hosting two ultra-dense objects in a close binary orbit, UCXBs are expected to play a crucial role in the new era of gravitational wave astronomy, as they will be among the loudest persistent sources \citep{Nelemans2018,Tauris2018,Chen2021a}. In addition, UCXBs are ideal targets to investigate and test some fundamental stages of binary evolution, such as the common-envelope phase, and are also unique laboratories to study accretion processes in hydrogen deficient environments \citep{Nelemans2009}.

We can distinguish three main UCXB evolutionary channels (see, e.g. \citealt{Nelemans2010b} for a more extended description): 

(i) The white dwarf (WD) channel. In this scenario, the binary system is formed by a black-hole/neutron-star and  a WD. The loss of angular momentum driven by gravitational wave emission shortens the binary period to a very compact configuration. At some point (generally at \Po\ of a few minutes), the WD overflows its Roche lobe and starts to transfer matter onto the more massive compact object, giving rise to a persistent UCXB. This also results in an expansion of the orbit. As \Po\ increases, the mass transfer rate decreases, making the systems less luminous. The UCXB eventually becomes a transient source as the accretion disc becomes not fully ionised and no longer stable \citep{Yungelson2002,Sengar2017,VanHaaften2012a}. 

(ii) The helium star channel. The companion star is a low-mass helium star, which is still burning helium in its core at the time of filling its Roche lobe. Mass transfer occurs as the orbit shrinks due to angular momentum loss via gravitational wave emission. A minimum \Po\ is reached ($\sim$10-60~min), when the helium core of the $\lesssim 0.2\,\Msun$ donor becomes semi-degenerate and its mass-radius relation inverts. This makes the companion star radius to increase with further mass loss \citep{Savonije1986,Iben1987,Yungelson2008,Nelemans2010b}. At this point, the orbit evolution reverses and the system expands as gravitational wave radiation can not counteract the effect of mass transfer. 

(iii) The evolved main-sequence star channel. The donor star is an evolved main-sequence star that started mass transfer near or just after the point of central hydrogen exhaustion. The orbit shrinks due to angular momentum loss via gravitational wave emission and magnetic braking, reaching orbital periods as short as $\sim$ 10~min. Then, the growing helium core becomes semi-degenerate and the system follows a similar path to that of the helium channel \citep[e.g.][]{Tutukov1985,Nelson1986,Nelson2003,Podsiadlowski2002,Sengar2017}.  

Interestingly, the final outcome of the above three channels is the same: an UCXB with a WD donor that is evolving towards longer orbital periods. The binary expansion is driven by conservation of angular momentum, as the less massive star (the WD) transfers mass to a heavier object. This slow evolutionary process could, in principle, increase \Po\ up to $\sim 120$~min in a Hubble time (\citealt{VanHaaften2012a,VanHaaften2012b}). Thus, while \Po $\lesssim 80$~min is a common observational diagnostic to separate UCXBs from the classic LMXB population (see e.g. \citealt{Nelemans2010} and \citealt{Heinke2013} for previous compilations of UCXBs), the former class might also include very old members with \Po\ up to 2 hours. These `long-period' UCXBs are, however, difficult to detect because of their much lower accretion rates and luminosities.

Globular clusters (GCs) have been proposed to host a significant fraction of the UCXB population.  The LMXB formation rate (per unit mass) in GCs is a hundred times higher than in the Galactic field (assuming that GCs account for $\sim0.01\%$ of the Galactic stellar mass; \citealt{Katz1975}). This overabundance is thought to be due to the fact that LMXB formation in GCs is strongly enhanced by dynamical encounters, such as tidal captures, direct collisions and exchange interactions \citep{Fabian1975,Hills1976,Verbunt1987,Rasio2000,Ivanova2005}. As a consequence, the populations of UCXBs formed in GCs through each of the above evolutionary pathways may be different from those in the Galactic field.  Therefore, the properties of the current (i.e. observable) populations of UCXBs in GCs and the Galactic field can be significantly different (e.g. distribution of \Po; \citealt{Zurek2009, Heinke2013}).

Measuring \Po\ in UCXB systems is often difficult because of the short time-scales and low luminosities involved. Thereby, the UCXB classification sometimes relies on a number of additional observational properties. The most direct ones, albeit not necessarily the most robust, are based on inferring the degenerate nature of the donor star from the chemical composition of the accreted material. These include: (i) the lack of hydrogen features in the optical spectra \citep{Nelemans2004,IntZand2008,Santisteban2019,ArmasPadilla2020,Stoop2021} together with the presence of helium and metallic lines (e.g. nitrogen and carbon; \citealt{Nelemans2004}; see also \citealt{Homer2002, Tudor2018} for UV examples); (ii) the presence of X-ray features attributed to overabundances (i.e., non-solar composition) in the accreted material or the local interstellar
medium (\citealt{Schulz2001,Juett2001,Krauss2007}, but see \citealt{Juett2005}) and enhanced fluorescent lines (e.g. O~\textsc{viii} Ly$\alpha$) due to X-ray reprocessing in an oxygen-rich accretion disc \citep[see e.g.][]{Madej2010,Schulz2010}; (iii) the properties of thermonuclear bursts, such as duration, recurrence time and radiated energy, which can provide hints on the composition of the accreted fuel (e.g. \citealt{Cumming2003, Falanga2008, Galloway2020}; see also \citealt{Juett2003b}). Additional diagnostics for identifying UCXBs are related to the small size of their accretion discs. First, UCXBs can be persistent sources at low X-ray luminosities (\lx$^{\rm pers}$~$\lesssim10^{36}\lum$) since smaller accretion discs can be entirely ionised at lower accretion rates \citep{Lasota2001,IntZand2007}. Secondly, given that the region of the disc emitting in the optical (via X-ray reprocessing) is also smaller, UCXBs show very low optical–to–X-ray flux ratios \citep{Paradijs1994}.

In this paper, we present a comprehensive catalogue of UCXBs, which we name \CAT, and is available online\footnote{\label{link}\url{https://research.iac.es/proyecto/compactos/UltraCompCAT/}}. Given the overall importance and very distinct features of the UCXB family, the main goal of this catalogue is to keep a complete and updated compilation of the current and future members of this LMXB subgroup, together with their observational properties. \CAT\ includes systems with a measured \Po\ shorter than 120 min and those without a solid \Po\ measurement, but which exhibit some of the UCXB-related features described above. The first version of the catalogue includes 49 LMXBs and is presented in section \ref{catalogue}. The most direct results derived from this compilation are presented in section \ref{results} and discussed in section \ref{sec:discussion}.

\begin{figure}
  \begin{center}
    \includegraphics[scale=1]{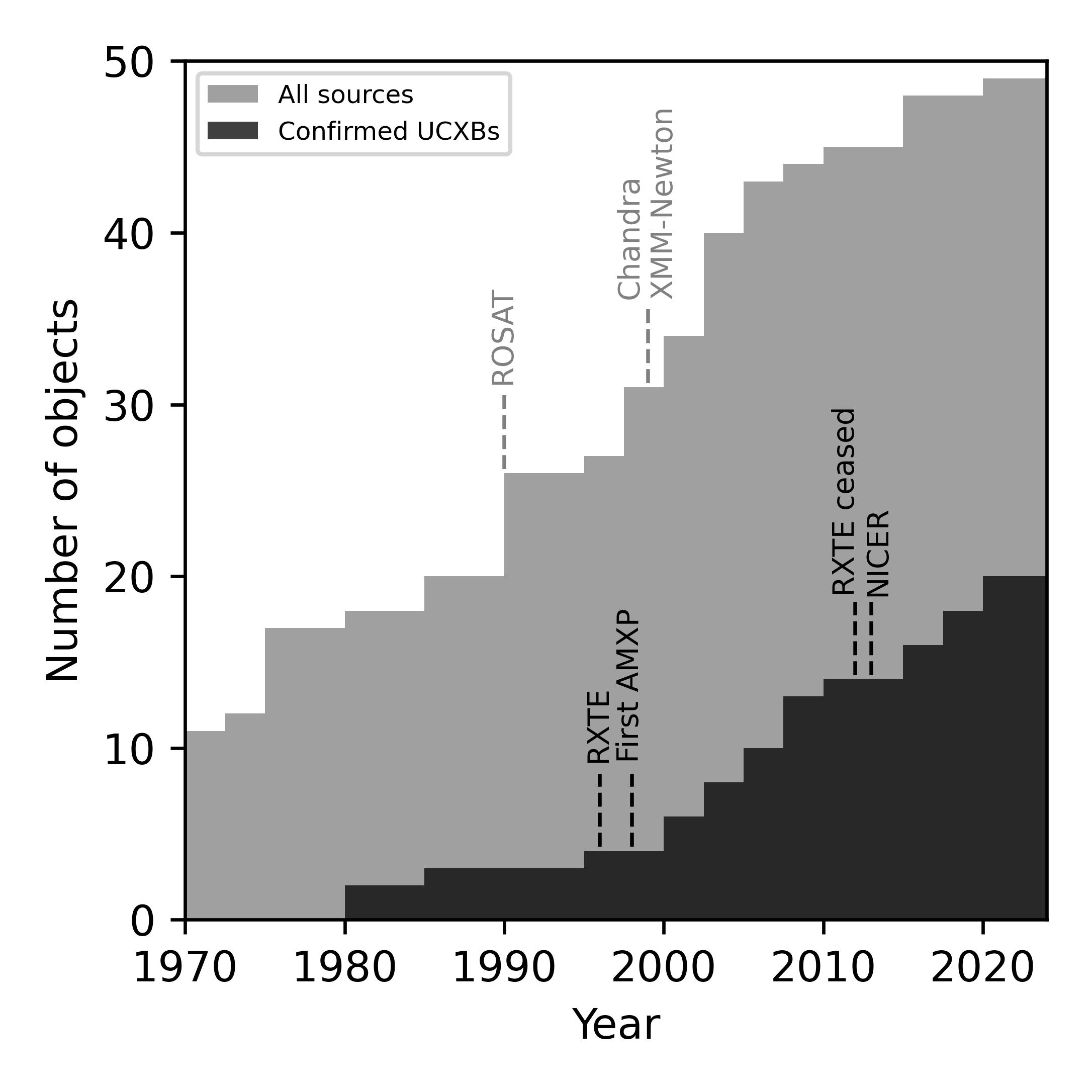}
    \caption[]{Cumulative histogram of ultra-compact and short orbital period LMXBs detected since the beginning of the X-ray astronomy era. The black bars represent the confirmed UCXBs.} 
    \label{fig:cumulative}
  \end{center}
\end{figure}

\section{The Catalogue}
\label{catalogue}

\subsection{The Sample}
We have searched the literature, collecting sources that can be divided in three main categories: ultra-compact, ultra-compact candidates and short orbital period LMXBs:
\paragraph{Ultra-compact X-ray binaries (UCXBs).}
These are LMXBs with orbital periods shorter than $\sim$80~min. This category includes both, LMXBs with confirmed orbital period determinations (e.g. derived from Doppler shifts in X-ray pulsars) and systems for which strong constraints, such as those inferred from periodic flux modulations, are available. The latter group might include systems where the orbital period is not known with high precision, as modulations often reflect periodicities that may deviate slightly from the orbital period (e.g. those resulting from super-humps are a few percent longer than \Po). A good example is 4U~0614+091, for which \citet{Shahbaz2008} reported a clear modulation and several candidate periods in the range of $\sim$ 40 -- 60 min, the strongest being $\sim$ 51 min. Other authors \citep[e.g.,][]{Nelemans2006,Hakala2011,Zhang2012,Baglio2014} have confirmed the presence of this modulation, reporting slightly different periods within the same range. Thus, while a precise value is still missing, we believe that the UCXB classification for this source is robust. Another example (albeit less extreme) is 4U~1820-303 for which clear X-ray and UV modulations at 11.41 and 11.5 min, respectively, have been found (\citealt{Stella1987} and \citealt{Wang2010}). This group currently includes 20 systems.

\begin{figure*}
  \begin{center}
    \includegraphics[scale=1]{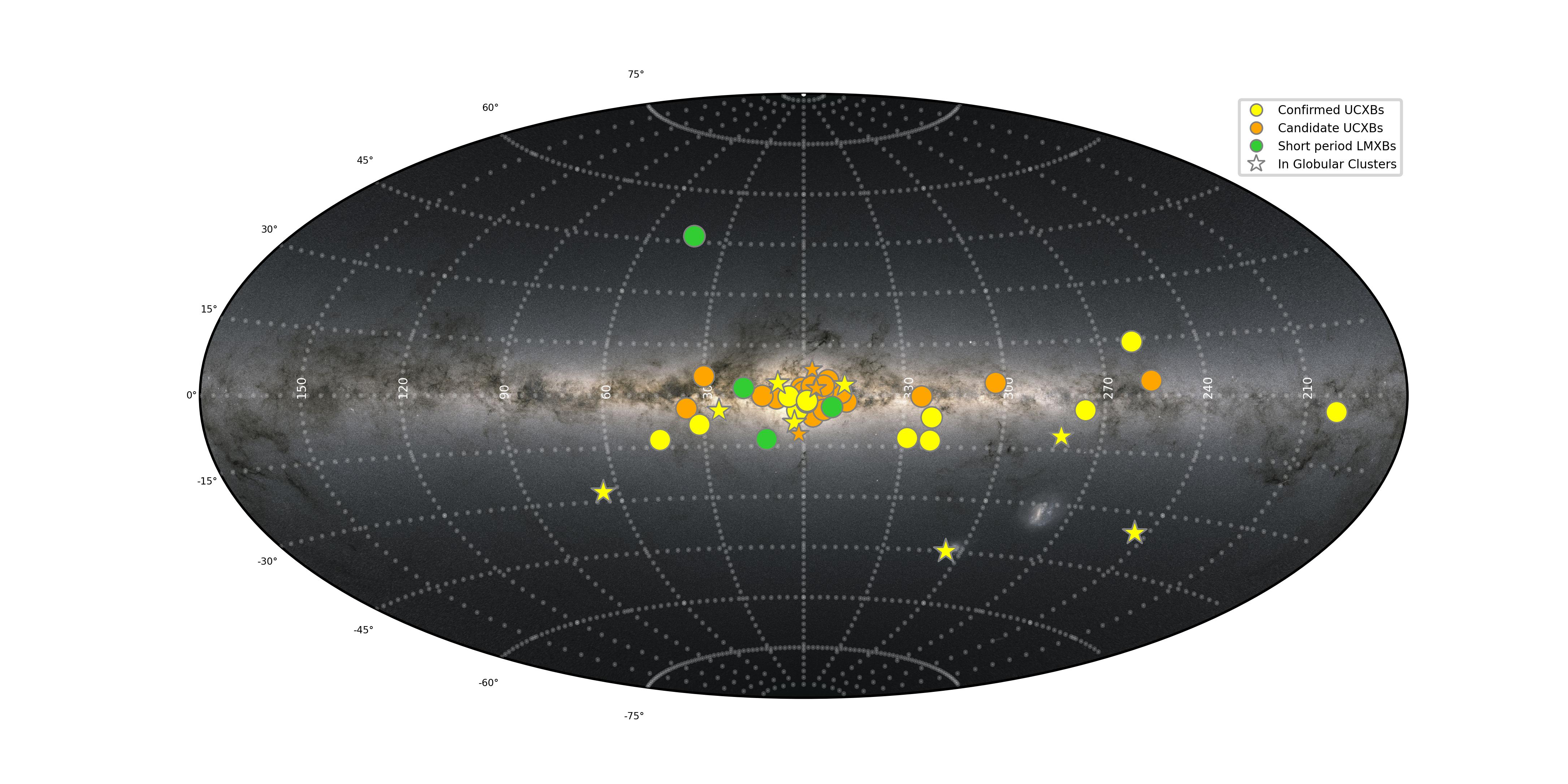}
    \includegraphics[scale=0.8]{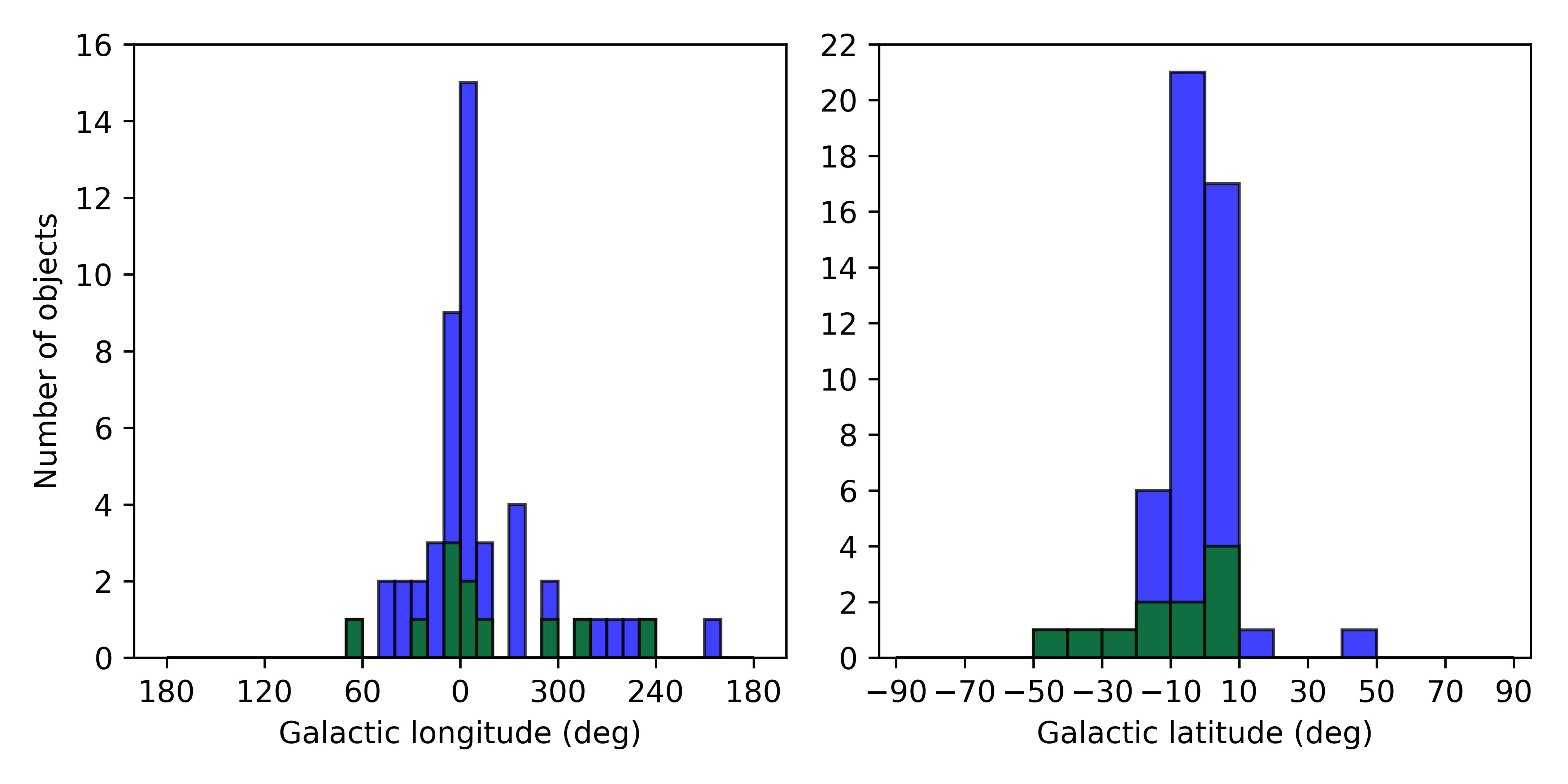}
    \caption[]{\textit{Top}: Galactic distribution for all the sources included in the catalogue: the 20 confirmed UCXBs (yellow), 25 candidates (orange) and four short-period LMXBs (green) in Galactic coordinates. Systems located in GCs are marked with stars rather than circles. [Background image credit: ESA/Gaia/DPAC]. \textit{Bottom}: Histogram of the distribution of all sources in Galactic longitude (left) and latitude (right). Blue is used for all systems (field + GC), and green for those located in GCs. A bin size of 10\,deg has been applied in both plots.}
    \label{fig:plane}
  \end{center}
\end{figure*} 

\paragraph{Ultra-compact X-ray binary candidates.}
LMXBs with suggestive UCXB properties (see above) are labelled as "candidates" (cUCXBs in Tables \ref{table:table1}, \ref{table:table2} and \ref{table:table3}). These systems might (or might not) have weak or contradictory constraints on the orbital period. This category currently counts 25 systems, including one object with no particularly suggestive UCXB properties but a proposed orbital period shorter than 80 min (NGC~6652~B,  43~min; \citealt{Deutsch2000}). We have also included in this category 4U~1728$-$34, which has a suggested \Po\ of 10.7~min (3$\sigma$ detection of an X-ray periodic signal) and thermonuclear bursts characteristic of hydrogen-poor fuel \citep{Galloway2010}. Alternatively, the observation of an infrared counterpart to a thermonuclear Type I X-ray burst points to \Po $>$1~hour \citep{Vincentelli2020}.

\paragraph{Short period LMXBs.}
This category is equivalent to that of UCXBs but for systems with \Po\ in the range of $\sim$ 80--120~min (spLMXBs in Tables \ref{table:table1}, \ref{table:table2} and \ref{table:table3}). The reason for having this category, currently including four sources, is twofold: (i) to retain old UCXBs that currently evolve towards long \Po $> 80$ min and (ii) include also close progenitors of UCXBs via the helium star and evolved main-sequence star channels (see e.g. \citealt{VanHaaften2012,ArmasPadilla2022}). We acknowledge that our choice for the \Po\ interval for this category is somewhat arbitrary. However, it should be mentioned that the LMXB with the shortest known orbital period not included in \CAT\ (i.e. with \Po\ $> 120$ min) is the accreting millisecond pulsar (AMXP) SAX~J1808.4-3658 with \Po=120.82~min. This is a very well studied source with observational properties that do not match those of UCXBs (e.g., it shows a classic hydrogen-rich optical spectrum; \citealt{Cornelisse2009}).

\subsection{UltraCompCAT}
The full catalogue is available online\footref{link} and has been built according to the criteria described here.

Table \ref{table:table1} presents the basic astrometric properties.
The column distribution is as follows:\par

\begin{description}

\item[(1)] ID number;
\item[(2)] Year of discovery;
\item[(3)] Source type: UCXB for confirmed UCXBs, cUCXB for candidate UCXBs, and spLMXB for short period LMXBs;
\item[(4)] Preferred name of the system (GC if located in one);\par
\item[(5-7)] Right ascension (RA) and declination (Dec) in equinox J2000. The accuracy in the astrometry and source of the coordinates are also reported;\par
\item[(8-9)] Galactic longitude ($\ell$) and latitude ($b$) in degrees;\par
\item[(10-11)] Estimated distance ($d$) and height above the Galactic plane ($z$) in kpc. We use the notation (H) and (He) for distances estimated using thermonuclear burst, assuming hydrogen-rich and pure helium, respectively. (G) distances from \gaia\ parallaxes. For systems located in GCs, we report the distance to their host GC (see Sect. \ref{sec:GalacDistr});\par
\item[(12)] References for the above parameters.

\end{description}

Table \ref{table:table2} summarizes the main properties for all the systems presented in Table \ref{table:table1}. The column distribution is:

\begin{description}
\item[(1)] ID number;
\item[(2)] Source type: UCXB for confirmed UCXBs, cUCXB for candidate UCXBs, and spLMXB for short period LMXBs;
\item[(3)] Preferred name of the system (GC if located in one);
\item[(4)] Type of accretion, persistent (P) or transient (T). LO is specified for those transients with very long outbursts, and QP for persistent systems with very short period of inactivity (see Sect. \ref{results});
\item[(5)] Confirmed nature of the accretor, black hole (BH) or neutron star (NS). If the NS is an accreting X-ray millisecond pulsar, AMXP is added;
\item[(6)] Reported \Po\ in min;
\item[(7)] Unabsorbed peak X-ray flux in $\flux$, standardised to the 2--10~keV band. To do so, we begin with the X-ray flux that has been published in the literature. We assume a power-law spectrum with a photon index $\Gamma$ = 2 (\citealt{Belloni2011}) and the published total neutral Galactic Hydrogen column density (\Nh), derived from direct X-ray spectral analysis (reported in column 9);
\item[(8)] Peak optical or IR apparent magnitude (and quiescence [q], if the system is transient). To document the original observed band, we provide its name in its original photometric system;
\item[(9)] Total neutral Galactic Hydrogen column density (\Nh), published in literature derived from direct X-ray spectral analysis;
\item[(10)] Optical Galactic extinction $E(B - V)$ reported in the literature for host GCs; 
\item[(11)] References for the above parameters
\end{description}

Table \ref{table:table3} shows the main multi-wavelength phenomenology supporting short orbital periods for the systems presented in Table \ref{table:table1}. The column distribution is:

\begin{description}
\item[(1)] ID number;
\item[(2)] Source type: UCXB for confirmed UCXBs, cUCXB for candidate UCXBs, and spLMXB for short period LMXBs;
\item[(3)] Preferred name of the system (GC if located in one);
\item[(4-6)] Detection of bursts and their types: short/normal burst (short-B), intermediate long burst (IB) or superburst (SB);
\item[(7-9)] Spectral information, in the X-ray, optical, and UV bands;
\item[(10-11)] Small accretion disc size indirect diagnostic properties: $L_{\rm o}/$\lx$<<$, for systems with low optical--to--X-ray luminosity ratio, and \lx$^\mathrm{Pers}<<$, for systems persistently accreting at very low X-ray luminosities;
\item[(12)] References for the above parameters.
\end{description}

\begin{figure}
  \begin{center}
    \includegraphics[scale=1]{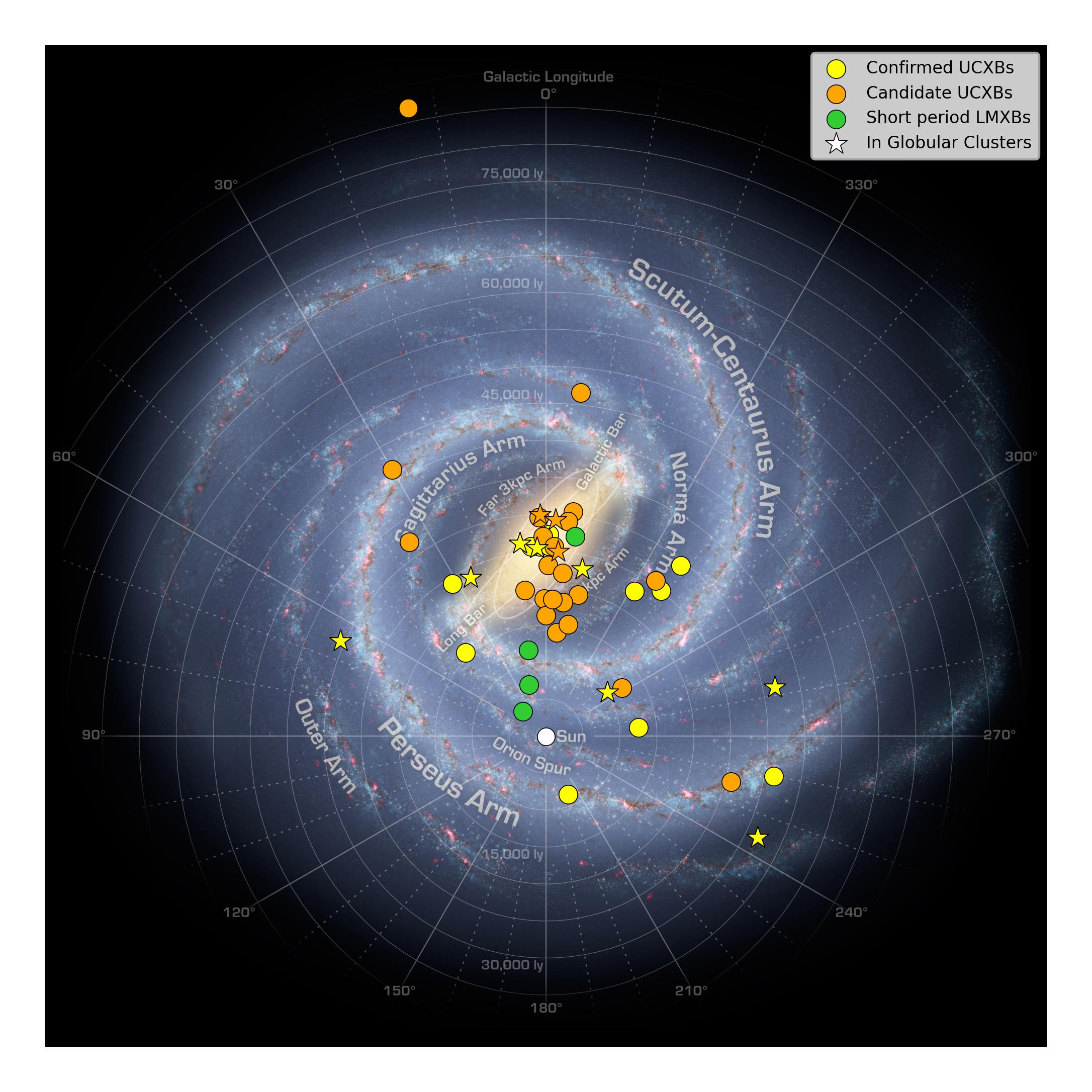}
    \caption[]{Galactic distribution of UCXBs and short-period LMXB systems with estimated distances following the same colours and symbols as in Fig.~\ref{fig:plane} [Background image credit: NASA/JPL-Caltech/R. Hurt (SSC/Caltech)]} 
    \label{fig:edgeon}
  \end{center}
\end{figure}

\begin{figure}
  \begin{center}
    \includegraphics[scale=1]{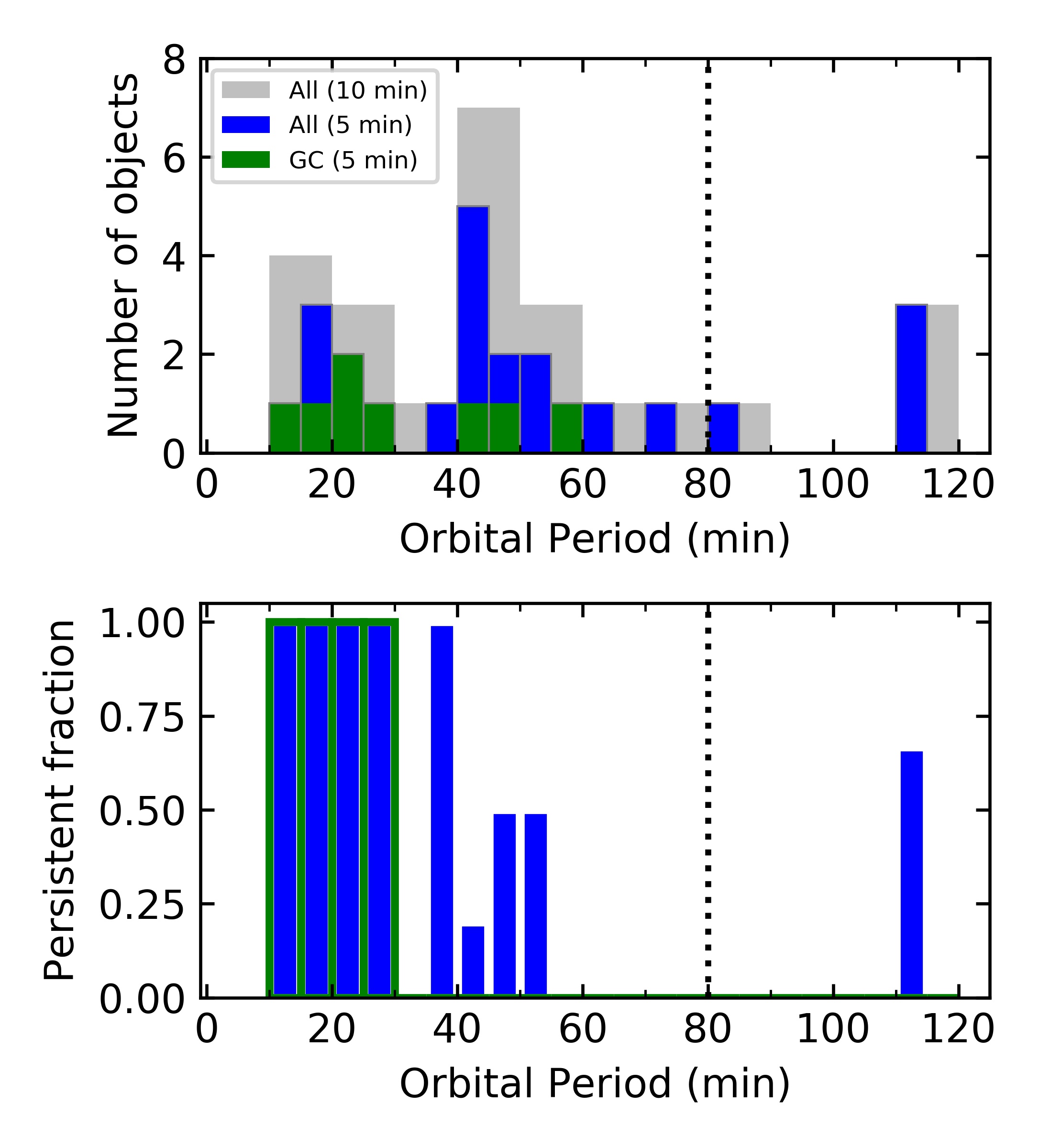}
    \caption[]{Top panel: Histogram of the orbital period of UCXBs and short-period LMXB systems, plotted in 10\,min bins (grey) and 5\,min bins [blue for all (field + GC) systems and green for those located in GCs]. The canonical 80\,min that defines bona fide UCXBs limit is shown in a dotted black line. Lower panel: Percentage of systems accreting persistently.} 
    \label{fig:porb}
  \end{center}
\end{figure}

\section{Results}\label{results}

\CAT\ currently includes 49 systems. A dozen of them were discovered at the  dawn of the X-ray astronomy in the 1970's. This number has steadily increased as new and more sensitive X-ray facilities were launched, such as the \rosat\ observatory in 1990, which was followed by \chan\ and \xmm\ missions in 1999 (see Fig. \ref{fig:cumulative}). It took 10 years for the first UCXB to be confirmed with the measurement of \Po=41.5~min in 4U 1626-67 using optical pulsations \citep{Middleditch1981}. The number of confirmed UCXBs increased significantly with the launch of missions with enhanced timing capabilities, such as \rxte\ (1996) and, more recently, \nicer\ (2017). This led to the discovery of AMXPs (\citealt{Wijnands1998,Chakrabarty1998a}; see \citealt{Campana2008,Patruno2021} for reviews), whose orbital periods can be efficiently measured with these facilities. To date, AMXPs account for half of the confirmed UCXB population (10 out of 20; see Table \ref{table:table2}). In addition, there are two other AMXPs in our sample that belong to the short period LMXB category (i.e., \Po\ in the range of 80 to 120~min).

Looking at the long-term behaviour, we find that 22 systems are transient (with five of them experiencing several years long outbursts), while the remaining 27 systems are persistent. The latter group includes four systems that can be classified as quasi-persistent, that is, sources that are active, but sporadically switch into short periods of low luminosity. An illustrative example is AX~J1754.2-2754, which has always been observed in active state, except for one occasion, when it was not detected for a brief period of time ($\lesssim$11~months; \citealt{Bassa2008,jonker2008,Degenaar2012d}).

The accretor is a NS in the vast majority of the systems (45 out of 49). This is known through the detection of thermonuclear bursts and/or pulsations, with 36 sources showing the former events. In particular, 18 sources have displayed intermediate-long bursts and four superbursts, with only two targets showing different burst durations compatible with the three categories (short, intermediate-long and superburst): 4U~0614+091 and SLX~1735-269 (see Table \ref{table:table3}). Finally, two of the targets that have not displayed NS signatures have been instead proposed to harbour BHs through indirect evidence. This is based on their position in the X-ray/radio, optical/X-ray and X-ray/photon-index planes (47~Tuc~X-9; \citealt{Miller-Jones2015} and IGR~J17285$-$2922; \citealt{Stoop2021}).
We note that this low fraction of BH systems is consistent with synthetic population models, which predict a very low number of BHs in UCXBs (less than 20\% \citealt{VanHaaften2013,Belczynski2004}). These would be mostly formed by accretion-induced collapse of accreting NSs \citep{Belczynski2004,Chen2023}, rather than a BH directly formed in the collapse of a massive star (but see \citealt{Qin2023}).

\subsection{Galactic distribution}\label{sec:GalacDistr}
The Galactic distribution of our sample is represented in Fig. \ref{fig:plane}. Half of them (24 out of the 49), are located in the central 20~deg in longitude (the Galactic bulge). The density of sources decreases drastically as we move away from the centre, with only a few targets scattered between 300 and 180~deg, and none between 70 and 180~deg (see also Fig. \ref{fig:edgeon}). In the same way, UCXBs and candidates are concentrated in the Galactic plane, with 92$\%$ of them comprised between -20 and 20~deg (77$\%$ between -10 and 10~deg; see Fig. \ref{fig:plane}). 

In Fig. \ref{fig:edgeon} we plot the objects with their estimated distances projected onto a face-on view of the Milky Way. About 35$\%$ of the sources are located within 8~kpc of the Sun, 50$\%$ between 8 and 10~kpc, and only 10$\%$ beyond 10~kpc (see also the distribution of distances in Fig. \ref{fig:dist-z}). We note that these numbers have to be taken with caution, since some distances are not well constrained. Half of the distances are derived from thermonuclear bursts assuming Eddington limit luminosity peaks. Here, we used helium fuel distances for the UCXBs and candidates, and hydrogen for the four short period LMXBs (i.e. we do not assume their UCXB nature). In addition, we note that the peak at 8--10~kpc includes nine systems, whose distances were assumed to be $\sim 8$ kpc ($\sim 10$~kpc in one case) based on their coordinates (i.e. consistent with the Galactic Centre) and neutral hydrogen equivalent column densities. For those systems located in GCs, we used the distance to the host GC. We used (meaningful) distances from \gaia\ parallaxes when the \gaia\ (candidate) counterpart was positively cross-matched to that of our X-ray binary system (e.g., those listed in the work of \citealt{Arnason2021}).

A total of 11 sources are located in GCs, eight confirmed UCXBs and three candidates (represented by yellow and orange stars, respectively, in Fig. \ref{fig:plane} and \ref{fig:edgeon}). Most of them correspond to those systems scattered away from the Galactic Bulge and Plane (see Fig. \ref{fig:plane}). Attending to their long-term variability, six of the sources are persistent systems (five in the case of the confirmed UCXBs; see Table \ref{table:table2}).

\subsection{Orbital period distribution}\label{sec:period}

The orbital period is known for 24 out of the 49 sources included in the catalogue. 
Fig. \ref{fig:porb} (top panel) shows the \Po\ distribution using bins of 5 and 10 min. The sources are grouped in two populations: one with \Po\ between 10 and 40 min (8 sources; peaking at $\sim$ 20 min), and a second group with \Po\ between 40 and 60 min (10 sources; peaking at 45 min). Only three systems are scattered within the range of 60 to 85 min, while none is found between 85 and 110 min. From there, the number of sources is observed to increase again, a trend that is known to be followed by classical LMXBs beyond the 120~min limit of \CAT\ (see e.g., \citealt{Bahramian2022}). It is worth noting that the ratio between the GC and field UCXB populations is very different in the two main peaks. While the 20-min group is dominated by systems in GCs (5 out of 8), the 45-min peak only includes three (out of 10), and thus is dominated by UCXBs formed in the field (i.e. Galactic Plane and Bulge).     

The lower panel in Fig. \ref{fig:porb} provides the percentage of systems that are persistent for each 5-min \Po\ bin. The objects within the $\sim$20\,min peak are all persistent. As a matter of fact, every UCXB below 40 min is persistent. This drops sharply to below $\sim$50$\%$ for \Po~$>40$~min, and then to zero for systems with \Po\ longer than 60 min. This number increases again to $\sim$60$\%$ for systems with \Po\ longer than 110~min. 

Out of the eight GC sources with a \Po\ measurement, five are persistent (those with \Po\ $<40$~min), while the remaining three sources of the 45-min group are transient. In the case of the 16 field systems, half of them are persistent: the three systems with \Po\ $<40$~min, three out of the seven systems in the 45-min peak, and two short-period LMXBs (with \Po$\sim$110~min).

\section{Discussion}\label{sec:discussion}
UCXBs are a distinctive subset of the LMXB family. In the past, this has given rise to several compilations  that list its members and basic characteristics (e.g. \citealt{IntZand2007,Nelemans2010,VanHaaften2012, Heinke2013}). In this paper, we have taken another step forward by creating \CAT. The UCXB properties compiled in the catalogue have allowed us to discuss their orbital period distribution, which is key to understand their origin and evolution, as well as their Galactic distribution.  

\subsection{UltraCompCAT: present and future}

\CAT\ is the first comprehensive catalogue of ultracompact and short orbital period X-ray binaries. It is available online and includes 49 sources, whose (known) properties have been carefully revised and listed. Among the hundreds of reported LMXBs, both transient and persistent, we have selected those with \Po\ or spectral properties compatible with the UCXB family. This has been done by thoroughly searching in the literature, aiming at providing accurate information that can be used in UCXB research. We cannot be totally certain that all the sources included in \CAT\ are bona fide UCXBs. First, we have extended the sample to systems with \Po\ $>80$ min, that we refer to as short-period LMXBs.  We find it is important to include this category since: (i) the UCXB nature of a given LMXB is not strictly defined by \Po, but by the nature of the donor. Systems with \Po\ $>80$ min may in fact be UCXBs that have evolved past the minimum \Po; (ii) even if they are not UCXBs, this subset of LMXBs can be useful for UCXB studies (since they might be UCXB progenitors; e.g. \citealt{ArmasPadilla2022}). In any case, only four out of the 49 systems belong to this \Po\ range. Secondly, we decided to follow an inclusive approach by building the UCXB-candidates group, although it could be the case that new \Po\ measurements end up disproving the UCXB nature of some of the current members. However, previous studies have shown that a significant fraction of the candidates are eventually confirmed \citep[e.g.][]{Zurek2009,Zhong2011,Strohmayer2018}. 

Likewise, \CAT\ is necessarily affected by observational biases related to the instrumentation employed,  cadence and completeness. Besides the ample theoretical support behind the idea that the vast majority of the Galactic UCXBs are yet to be discovered, it is also possible that some known LMXBs are unclassified UCXBs. Similarly, there is a large number of faint, unclassified X-ray sources detected by the most sensitive X-ray surveys \citep[e.g.][]{Webb2020,Bahramian2021}. While the dominant population is expected to be (magnetic) accreting WDs, the large number of Galactic UCXBs predicted to exist by theoretical studies \citep{Belczynski2004,VanHaaften2013} suggest the UCXB nature of some of these faint sources. 

Despite the above biases, \CAT\ represents the current observational view of UCXBs, and the most complete ever compiled.  \CAT\ will be continuously updated, growing in size as more systems are discovered and characterised, and more orbital periods are measured.

\subsection{On the different populations of UCXBs}

The Galactic distribution of UCXBs (see Fig. \ref{fig:plane}) shows an accumulation of objects in the Galactic bulge (although several distances are not well constrained) and along the disc plane, with only a few scattered sources placed further away (located mostly in GCs). This is in agreement with UCXBs being mostly Population-II stars \citep{Tauris2006}. About a quarter of our sources are located in GCs: eight of the 20 confirmed UCXBs and three of the candidates. This is not unexpected since UCXBs are predicted to be overproduced in these dense conglomerations of stars via dynamical collisions. As a matter of fact, UCXBs formed in GCs have even been suggested to be one of the dominant populations of X-ray binaries \citep{Verbunt1987,Bildsten2004,Ivanova2008}.

The histogram representing the \Po\ distribution (Fig. \ref{fig:porb}) shows three main groups, which might be related to different UCXBs populations. One group with \Po $\sim$ 20 min (10 to 40 min), a second one with \Po $\sim$ 45 min (40 to 60 min), and a third, less populated, with periods longer than 110 min. According to binary synthesis models, the population of UCXBs in the Galaxy is expected to be $\sim~0.1-1\times10^{5}$. Most of them are predicted to have already evolved to \Po\ longer than $\sim$60~min, and thus characterised by low accretion rates. Taking this into account, together with the disc instability model, only a few tens of UCXBs are expected to have shown activity during the X-ray astronomy era \citep{Zhu2012,Belczynski2004}. However, systems formed in GCs via dynamical interactions (i.e. at any time) may modify this picture.

\citet{VanHaaften2013} explored the formation of UCXBs in the Galactic bulge (i.e. the oldest Galactic field population). They predict a peak of sources below 30 min dominated by persistent systems formed through the WD channel, most of them with He or C/O WD donors. Contrastingly, the helium burning channel is expected to contribute (at most) with a few transients with longer \Po\ (60--80 min). Hence, the observed 20-min peak formed by persistent systems (Fig. \ref{fig:porb}) could in principle be associated with the WD channel. However, Fig. \ref{fig:porb} shows that it is dominated by GC systems (75 \%), while very short \Po\ ($\lesssim$ 40 min) UCXBs formed in the field are scarce. The histogram also includes three transient field UCXBs with \Po\ in the range of 60-80~min, that could be identified with the helium burning channel according to \citet{VanHaaften2013}, but this work does not specifically account for the observed peak at 40--60 min.

In a different study, \citet{Zhu2012} predict that systems formed via the helium star channel in the Galactic field might be observed over a wider range in \Po\ (between $\sim1-80$~min), peaking at $\sim$40~min. These systems would be persistent, while the observed 40--60~min population is mostly transient. In this model, transient UCXBs would have longer periods (33$<$\Po$<$130~min), but only a few (i.e. many less than persistent ones) would be detected. This does not seem to be consistent with the current (i.e. observed) Galactic field populations, in which half of UCXBs are transient sources. The evolved main-sequence star channel is predicted to produce a much lower number of UCXBs (e.g., only 0.3 \% according to \citealt{VanHaaften2013}), and thus its contribution to the observable population should be negligible. 

In general, UCXBs population synthesis models, which are mostly focused on systems in the Galactic field, predict that the vast majority of the sources should have evolved towards relatively long orbital periods. Therefore, they would be transients characterised by low accretion rates (and thus long recurrence times). However, our study shows that the transient UCXB population (i.e. those that have displayed at least one outburst in $\sim$ 60 years) is not as low as predicted. As a matter of fact, transients account for almost half of the sample (22 out of 49 systems, 9 out of the 20 confirmed UCXBs). This becomes even more striking if we only consider UCXBs in the Galactic field (19 persistent and 19 transients). On the other hand, the population of UCXBs and candidates in GCs is formed by 11 sources, most of them persistent with \Po\ $<$ 30 min (five out of the eight with a \Po\ measurement). In fact, there is a dearth of (longer period) transients formed in GCs. This might be consistent with most of them being produced by dynamical interactions (i.e. at any time) via the WD evolutionary channel, as suggested in \citet{Heinke2013}.

To elucidate the evolutionary channel of a given UCXB based on observational constraints is challenging. Optical, X-ray and UV spectroscopy, together with the properties of type I X-ray bursts can be used to infer the abundances of the accreted material. For instance, \citet{Cumming2003} propose that 4U~1820$-$303 has a helium rich companion based on its type I X-ray bursts. However, tight constraints on more than one element would be required to single out the evolutionary channel (see e.g. \citealt{Nelemans2010b}). In addition, one might also consider the observed accretion rate (i.e. luminosity) for a given \Po\ together with the disc instability model (i.e. persistent or transient behaviour). Based on this, the three persistent systems of the 45-min group may have formed through the helium star channel \citep{Nelemans2010b,Heinke2013}. Nevertheless, its worth noting that there are several cases where it is difficult to reconcile all the above into a single picture (e.g. 4U~0614+091 and 2S~0918-549; see e.g. \citealt{Kuulkers2010,Heinke2013,ArmasPadilla2020}).

\section{Conclusions}\label{Conclusions}
We have presented the first comprehensive catalogue of ultracompact and short orbital period X-ray binaries, which currently includes 49 sources. This compilation is available online\footref{link} and includes a carefully revised collection of observational properties for each source. \CAT\ will grow as more systems are discovered and more orbital periods are measured. Thus, we expect that it will become a powerful tool for the study of this key subgroup of LMXBs, particularly in the new era of gravitational wave astrophysics, since UCXBs are expected to be among the loudest persistent sources. We have shown that the orbital period distribution of the current sample presents two main groups. One mainly formed by persistent
systems in GCs with orbital periods shorter than 30 min, and a second one with transient objects (70 \%), periods in the range of 40 to 60 min, and dominated by UCXBs in the Galactic field. These two groups might be the result of two different evolutionary channels that dominate in the GC and field populations. Future observations of known and newly discovered UCXBs will allow \CAT\ to grow and hopefully shed light on these open problems.

\begin{acknowledgements}
    This work is supported by the Spanish Ministry of Science under grants PID2020--120323GB--I00, PID2021--124879NB-I00, and EUR2021--122010. We acknowledge support from the Consejería de Economía, Conocimiento y Empleo del Gobierno de Canarias and the European Regional Development Fund (ERDF) under grant with reference ProID2021010132.
\end{acknowledgements}

%
%

\bibliographystyle{aa} 
\bibliography{UltraCompCAT.bbl}{} %

\begin{appendix}

\begin{figure*}
\section{Histogram}
  \begin{center}
    \includegraphics[width=\textwidth]{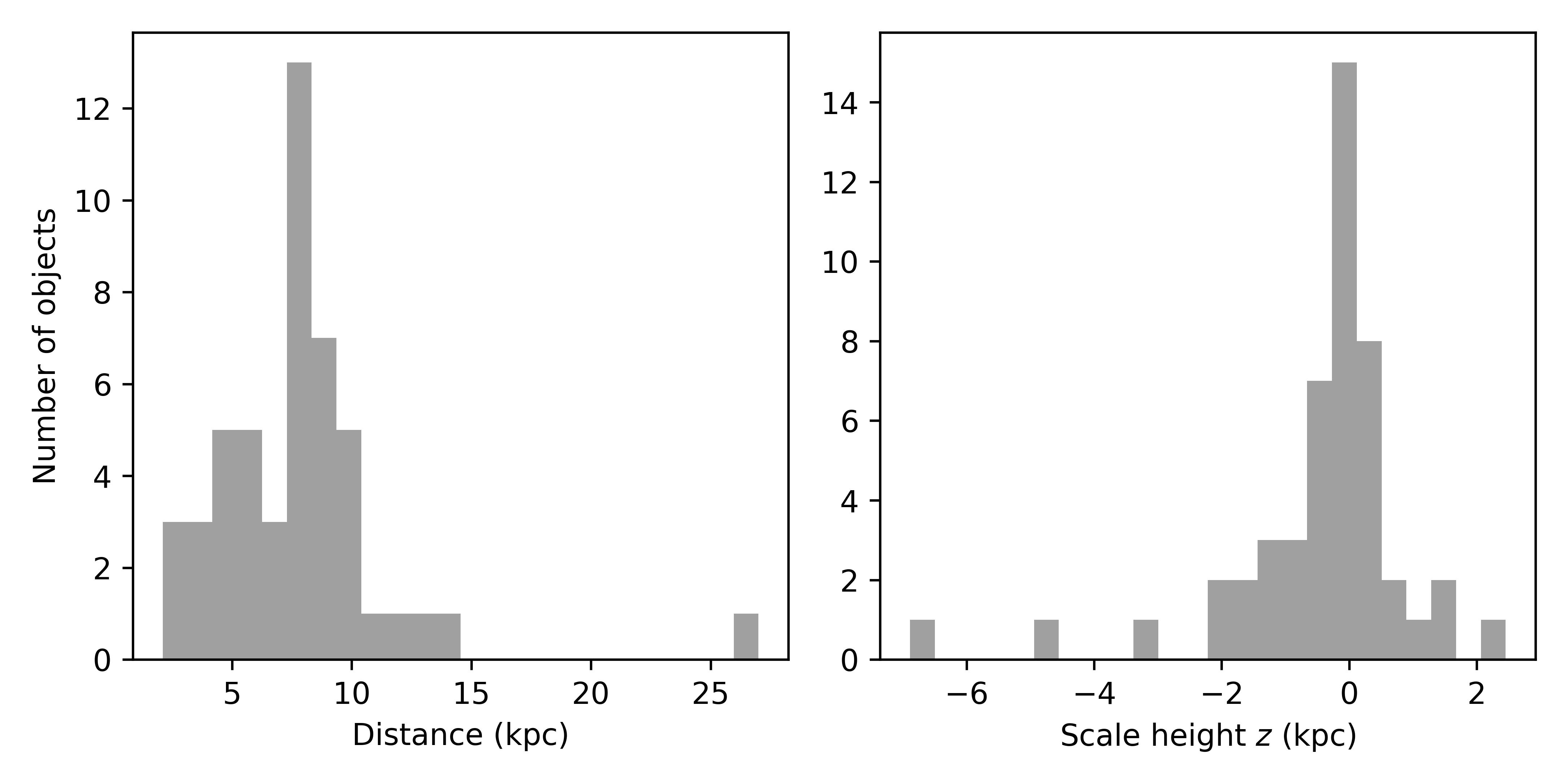}
    \caption[]{Histogram of the distribution of all the objects with known distances (left) and their scale height (\textit{z}) from the Galactic plane (right).} 
    \label{fig:dist-z}
  \end{center}
\end{figure*}

\clearpage

\onecolumn
\begin{landscape}\begin{ThreePartTable}
\section{Tables}
\begin{TableNotes}\scriptsize\item \textbf{Notes. } \tablefoottext{a}{In column 7, the letter preceding the error value indicate the source of the coordinates: radio (r), optical (o), infrared (i), ultraviolet (uv), X-rays (x), Gaia (G) or a combination of them. If RA and Dec errors are different we give both.}\tablefoottext{b}{ We use the notation (H) and (He) for distances estimated using thermonuclear bursts, assuming hydrogen-rich and pure helium, respectively. (G) distances from \gaia\ parallaxes. For systems located in GCs, we report the distance to their host GC (see Sec. \ref{sec:GalacDistr}). In the references column (12), “\textit{y}” refers to discory year, “\textit{c}” to coordinates, and “\textit{d}” to distance.}\end{TableNotes}\setlength{\tabcolsep}{2.pt}\setlength{\LTcapwidth}{\textwidth}\scriptsize\centering\begin{longtable}{llllllllllll}\caption{\label{table:table1} Astrometry and location.}\\\hline\hline\\[-0.15cm](1) & (2) & (3) &(4) &(5)&(6)&(7)&(8) &(9)&(10)& (11) & (12)\\ID & Type & Year & Name (GC) & RA & Dec & Error \tablefootmark{a}  & $\ell$ & $b$ & $d$ \tablefootmark{b}  & $z$ & References \\ &  &  &  & (h m s) & (\degr~\arcmin~\arcsec ) & (\arcsec ) & (\degr) & (\degr) & (kpc) & (kpc) &  \\\endfirsthead\caption{continued.}\\\hline\hline\\[-0.15cm](1) & (2) & (3) &(4) &(5)&(6)&(7)&(8) &(9)&(10)& (11) & (12)\\ID & Type & Year & Name (GC) & RA & Dec & Error \tablefootmark{a}  & $\ell$ & $b$ & $d$ \tablefootmark{b}  & $z$ & References \\ &  &  &  & (h m s) & (\degr~\arcmin~\arcsec ) & (\arcsec ) & (\degr) & (\degr) & (kpc) & (kpc) &  \\\hline\\[-0.15cm]\endhead\hline\insertTableNotes\endfoot\hline\\\endlastfoot\hline\\[-0.15cm]

1 & UCXB & 1972 & 4U 1820-303 (NGC 6624) & 18:23:40.50 & -30:21:40.09 & (r) 0.0004,0.005 & 2.788211 & -7.913491 & 8.019 & -1.101 & \makecell[tl]{\citet{Giacconi1972}$^{y}$, \citet{DiazTrigo2017}$^{c}$\\ \citet{Baumgardt2021}$^{d}$} \\
2 & UCXB & 1972 & 4U 0513-40 (NGC 1851) & 05:14:06.40 & -40:02:38 & (o) 0.7 & 244.509804 & -35.036749 & 11.951 & -6.889 & \makecell[tl]{\citet{Forman1976b}$^{y}$, \citet{Homer2001}$^{c}$\\ \citet{Baumgardt2021}$^{d}$} \\
3 & UCXB & 1974 & 2S 0918-549 & 09:20:26.95 & -55:12:24.7 & (x) 0.6 & 275.852936 & -3.844515 & 3.0 (H), 3.9 (He) & -0.261 & \makecell[tl]{\citet{Giacconi1974}$^{y}$, \citet{Juett2003b}$^{c}$\\ \citet{Galloway2020}$^{d}$} \\
4 & UCXB & 1972 & 4U 1543-624 & 15:47:54.69 & -62:34:05.4 & (x) 0.6 & 321.756967 & -6.336329 & 9.2 (He) & -1.015 & \makecell[tl]{\citet{Giacconi1972}$^{y}$, \citet{Juett2003b}$^{c}$\\ \citet{Serino2018}$^{d}$} \\
5 & UCXB & 1975-1976 & 4U 1850-087 (NGC 6712) & 18:53:04.88 & -08:42:20.0 & (r) 0.4 & 25.355515 & -4.319866 & 7.382 & -0.557 & \makecell[tl]{\citet{Seward1976,Swank1976}$^{y}$, \citet{Lehto1990}$^{c}$\\ \citet{Baumgardt2021}$^{d}$} \\
6 & UCXB & 2000 & M15 X-2 (M15/NGC 7078) & 21:29:58.14 & +12:10:01.52 & (uv) & 65.012156 & -27.311977 & 10.709 & -4.910 & \makecell[tl]{\citet{White2001}$^{y}$, \citet{Haurberg2010}$^{c}$\\ \citet{Baumgardt2021}$^{d}$} \\
7 & UCXB & 1981 & 47 Tuc X-9 (47 Tuc/NGC 104) & 00:24:04.25 & -72:04:58.02 & (x) 0.001, 0.01 & 305.89703 & -44.887576 & 4.521 & -3.176 & \makecell[tl]{\citet{Grindlay1981}$^{y}$, \citet{Heinke2005}$^{c}$\\ \citet{Baumgardt2021}$^{d}$} \\
8 & UCXB & 2006 & IGR J17062-6143 & 17:06:16.3 & -61:42:40.5 & (uv) 0.3 & 328.719432 & -12.398127 & 7.3 (He) & -1.567 & \makecell[tl]{\citet{Churazov2007}$^{y}$, \citet{Ricci2008}$^{c}$\\ \citet{Keek2017}$^{d}$} \\
9 & UCXB & 2003 & XTE J1807-294 & 18:06:59.8 & -29:24:30 & (x) 1 & 1.935298 & -4.272579 & assumed 8 & -0.596 & \makecell[tl]{\citet{Markwardt2003a}$^{y}$, \citet{Markwardt2003}$^{c}$\\ \citet{Campana2003}$^{d}$} \\
10 & UCXB & 1972 & 4U 1626-67 & 16:32:16.79 & -67:27:39.3 & (x) & 321.788351 & -13.091711 & ~8 & -1.812 & \makecell[tl]{\citet{Giacconi1972}$^{y}$, \citet{Lin2012a}$^{c}$\\ \citet{Chakrabarty1998}$^{d}$} \\
11 & UCXB & 2002 & XTE J1751-305 & 17:51:13.49 & -30:37:23.4 & (x) 0.05, 0.6 & 359.181912 & -1.912008 & assumed 8.5 & -0.284 & \makecell[tl]{\citet{Markwardt2002}$^{y}$, \citet{Markwardt2002}$^{c}$\\ \citet{Gierlinski2005}$^{d}$} \\
12 & UCXB & 2002 & XTE J0929-314 & 09:29:20.19 & -31:23:03.2 & (o) 0.1 & 260.105098 & 14.214563 & assumed 10 & 2.456 & \makecell[tl]{\citet{Remillard2002a}$^{y}$, \citet{Giles2005}$^{c}$\\ \citet{Campana2005}$^{d}$} \\
13 & UCXB & 2016 & MAXI J0911-655 (NGC 2808) & 09:12:02.43 & -64:52:06.4 & (x) 0.6 & 282.195949 & -11.256788 & 10.060 & -1.952 & \makecell[tl]{\citet{Serino2016}$^{y}$, \citet{Homan2016a}$^{c}$\\ \citet{Baumgardt2021}$^{d}$} \\
14 & UCXB & 2017 & IGR J16597-3704 (NGC 6256) & 16:59:32.90 & -37:07:14.28 & (r) 0.005,0.22 & 347.792766 & 3.306254 & 7.242 & 0.415 & \makecell[tl]{\citet{Bozzo2017}$^{y}$, \citet{Tetarenko2018a}$^{c}$\\ \citet{Baumgardt2021}$^{d}$} \\
15 & UCXB & 1972 & 4U 1916-053 & 19:18:47.87 & -05:14:17.09 & (x) 0.6 & 31.357765 & -8.463592 & 5.8(H), 7.6(He) & -1.119 & \makecell[tl]{\citet{Giacconi1972}$^{y}$, \citet{Iaria2006}$^{c}$\\ \citet{Galloway2020}$^{d}$} \\
16 & UCXB & 1972 & 4U 0614+091 & 06:17:07.35 & +09:08:13.4 & (r) 0.1 & 200.877357 & -3.363284 & 1.99(H), 2.59(He) & -0.153 & \makecell[tl]{\citet{Giacconi1972}$^{y}$, \citet{Migliari2010}$^{c}$\\ \citet{Galloway2020}$^{d}$} \\
17 & UCXB & 2007 & Swift J1756.9-2508 & 17:56:57.43 & -25:06:27.4 & (x) 0.5 & 4.580765 & -0.212396 & assumed 8 & -0.030 & \makecell[tl]{\citet{Krimm2007}$^{y}$, \citet{Sanna2018a}$^{c}$\\ \citet{Krimm2007}$^{d}$} \\
18 & UCXB & 2009 & NGC 6440 X-2 (NGC 6440) & 17:48:52.76 & -20:21:24.0 & (x) 0.02,0.1 & 7.731804 & 3.802875 & 8.248 & 0.544 & \makecell[tl]{\citet{Heinke2010}$^{y}$, \citet{Heinke2010}$^{c}$\\ \citet{Baumgardt2021}$^{d}$} \\
19 & UCXB & 2015 & MAXI J1957+032 & 19:56:39.11 & +03:26:43.66 & (x) 0.6 & 43.645509 & -12.825722 & 5 & -1.110 & \makecell[tl]{\citet{Negoro2015b,Cherepashchuk2015}$^{y}$, \citet{Chakrabarty2016}$^{c}$\\ \citet{MataSanchez2017a,Ravi2017}$^{d}$} \\
20 & UCXB & 2012 & IGR J17494-3030 & 17:49:23.62 & -30:29:59.0 & (x) 0.6 & 359.086011 & -1.510211 & assumed 8 & -0.211 & \makecell[tl]{\citet{Boissay2012}$^{y}$, \citet{Chakrabarty2020}$^{c}$\\ \citet{ArmasPadilla2013c}$^{d}$} \\
21 & spLMXB & 2005 & HETE J1900.1–2455 & 19:00:09.77 & -24:54:04 & (r) 0.1,0.1 & 11.324799 & -12.868715 & 2.9(H), 3.8(He) & -0.846 & \makecell[tl]{\citet{Vanderspek2005}$^{y}$, \citet{Rupen2005}$^{c}$\\ \citet{Galloway2020}$^{d}$} \\
22 & spLMXB & 1990 & 1E 1603.6+2600 & 16:05:45.87 & +25:51:45.24 & (G) 0.0001, 0.0002 & 42.750438 & 46.77836 & 2.1 Gaia & 1.530 & \makecell[tl]{\citet{Gioia1990}$^{y}$, \citet{Gaia2020,Morris1990}$^{c}$\\ \citet{Arnason2021}$^{d}$} \\
23 & spLMXB & 2004 & IGR J17379–3747 & 17:37:58.84 & -37:46:18.35 & (r) 0.02, 0.07 & 351.648136 & -3.305259 & assumed 8.5 & -0.490 & \makecell[tl]{\citet{Chelovekov2006}$^{y}$, \citet{VanDenEijnden2021}$^{c}$\\ \citet{Chelovekov2006}$^{d}$} \\
24 & spLMXB & 1972 & 4U 1812-12 & 18:15:06.15 & -12:05:46.7 & (o) 0.19, 0.21 & 18.032721 & 2.397809 & 2.3(H)-3(He) & 0.096 & \makecell[tl]{\citet{Giacconi1972}$^{y}$, \citet{Bassa2006}$^{c}$\\ \citet{Galloway2020}$^{d}$} \\
25 & cUCXB & 1972 & 4U 1728-34 & 17:31:57.68 & -33:50:01.54 & (r) 0.0003, 0.005 & 354.303524 & -0.150417 & 3.3(H), 4.4 (He); & -0.012 & \makecell[tl]{\citet{Giacconi1972}$^{y}$, \citet{DiazTrigo2017}$^{c}$\\ \citet{Galloway2020}$^{d}$} \\
26 & cUCXB & 1998 & NGC 6652 B (NGC 6652) & 18:35:44.54 & -32:59:38.97 & (x) 0.009, 0.09 & 1.529081 & -11.37483 & 9.464 & -1.874 & \makecell[tl]{\citet{Deutsch1998,Heinke2001}$^{y}$, \citet{Heinke2001}$^{c}$\\ \citet{Baumgardt2021}$^{d}$} \\
27 & cUCXB & 1972 & 4U 1822-00 & 18:25:22.02 & -00:00:43.0 & (x) 0.6 & 29.939078 & 5.792617 & 13 & 1.312 & \makecell[tl]{\citet{Giacconi1972}$^{y}$, \citet{Juett2005}$^{c}$\\ \citet{Asai2016}$^{d}$} \\
28 & cUCXB & 1976 & 4U 1905+000 & 19:08:26.97 & +00:10:07.7 & (o) & 35.024458 & -3.707147 & 7.5 (H), 10 (He) & -0.647 & \makecell[tl]{\citet{Seward1976,Villa1976}$^{y}$, \citet{Gottwald1991}$^{c}$\\ \citet{Jonker2004}$^{d}$} \\
29 & cUCXB & 1990 & RX J1709.5-2639 (NGC 6293) & 17:09:30.4 & -26:39:19.9 & (x) 0.6 & 357.472538 & 7.911726 & 9.192 & 1.266 & \makecell[tl]{\citet{Voges1999}$^{y}$, \citet{Jonker2003}$^{c}$\\ \citet{Baumgardt2021}$^{d}$} \\
30 & cUCXB & 1999 & AX J1754.2-2754 & 17:54:14.5 & -27:54:35.6 & (x) 0.45 & 1.853974 & -1.099765 & 6.6(H),9.2(He) & -0.177 & \makecell[tl]{\citet{Sakano2002}$^{y}$, \citet{Bassa2008}$^{c}$\\ \citet{Chelovekov2007}$^{d}$} \\
31 & cUCXB & 1990 & 1RXS J171824.2-402934 & 17:18:24.14 & -40:29:33.04 & (x) 0.6 & 347.276837 & -1.651634 & 4.7(H),6.1(He) & -0.176 & \makecell[tl]{\citet{Voges1999}$^{y}$, \citet{IntZand2005}$^{c}$\\ \citet{Galloway2020}$^{d}$} \\
32 & cUCXB & 1990 & 1RXH J173523.7-354013 & 17:35:23.75 & -35:40:16.1 & (uv) 0.56 & 353.144504 & -1.741226 & 6.2(H),9.5(He) & -0.289 & \makecell[tl]{\citet{Voges1999}$^{y}$, \citet{Israel2008}$^{c}$\\ \citet{Degenaar2010b}$^{d}$} \\
33 & cUCXB & 2003 & XMMU J174716.1-281048 & 17:47:16.16 & -28:10:48.0 & (x) 0.5 & 0.834266 & 0.083421 & <5.3 (H),<8.4 (He) & 0.012 & \makecell[tl]{\citet{Sidoli2003}$^{y}$, \citet{Degenaar2007}$^{c}$\\ \citet{degenaar2011_burstxmmsource}$^{d}$} \\
34 & cUCXB & 1996-1997 & SAX J1806.5-2215 & 18:06:32.17 & -22:14:17.32 & (i) 0.03, 0.05 & 8.159359 & -0.699213 & <4.8 (H),<6.2 (He) & -0.076 & \makecell[tl]{\citet{intzand1999}$^{y}$, \citet{Kaur2017}$^{c}$\\ \citet{Galloway2020}$^{d}$} \\
35 & cUCXB & 1998 & AX J1538.3-5541 & 15:38:14.00 & -55:42:13.6 & (x) 0.8 & 324.945782 & -0.130139 & assumed 8 & -0.018 & \makecell[tl]{\citet{Sugizaki2001,Degenaar2012d}$^{y}$, \citet{Anderson2014}$^{c}$\\ \citet{Degenaar2012d}$^{d}$} \\
36 & cUCXB & 2020 & Swift J0840.7-3516 & 08:40:40.94 & -35:16:25.1 & (o) 0.2 & 256.281083 & 4.017002 & assumed 8 & 0.560 & \makecell[tl]{\citet{Evans2020}$^{y}$, \citet{Malesani2020}$^{c}$\\ \citet{CotiZelati2021}$^{d}$} \\
37 & cUCXB & 1976 & 1A 1246-588 & 12:49:39.36 & -59:05:14.68 & (o) 0.06, 0.05 & 302.702563 & 3.783544 & 3.0 (H), 3.8 (He) & 0.251 & \makecell[tl]{\citet{Seward1976b}$^{y}$, \citet{Bassa2006}$^{c}$\\ \citet{Galloway2020}$^{d}$} \\
38 & cUCXB & 1999 & SAX J1712.6-3739 & 17:12:36.77 & -37:38:41.0 & (x) 0.6 & 348.93894 & 0.919334 & 3.7(H), 4.8(He) & 0.077 & \makecell[tl]{\citet{Intzand1999b}$^{y}$, \citet{Wiersema2009}$^{c}$\\ \citet{Galloway2020}$^{d}$} \\
39 & cUCXB & 1972 & 1RXS J170854.4-321857 & 17:08:54.27 & -32:19:57.13 & (x) 0.6 & 352.780023 & 4.671897 & <4.4(H), <5.7(He) & 0.464 & \makecell[tl]{\citet{Giacconi1972}$^{y}$, \citet{IntZand2005}$^{c}$\\ \citet{Galloway2020}$^{d}$} \\
40 & cUCXB & 1975 & 4U 1722-30 (Terzan 2) & 17:27:33.15 & -30:48:07.8 & (x) 0.6 & 356.319615 & 2.298083 & 7.753 & 0.313 & \makecell[tl]{\citet{Swank1977}$^{y}$, \citet{Revnivtsev2002b}$^{c}$\\ \citet{Baumgardt2021}$^{d}$} \\
41 & cUCXB & 2003 & IGR J17254-3257 & 17:25:24.8 & -32:57:15 & (x) 2 & 354.279061 & 1.47467 & <14.5 (He) & 0.373 & \makecell[tl]{\citet{Walter2004}$^{y}$, \citet{Chenevez2007}$^{c}$\\ \citet{Chenevez2007}$^{d}$} \\
42 & cUCXB & 1985 & SLX 1735-269 & 17:38:17.12 & -26:59:38.6 & (x) 0.6 & 0.796121 & 2.40017 & 4.5(H), 5.8(He) & 0.243 & \makecell[tl]{\citet{Skinner1987}$^{y}$, \citet{Wilson2003}$^{c}$\\ \citet{Galloway2020}$^{d}$} \\
43 & cUCXB & 1985 & SLX 1737-282 & 17:40:42.83 & -28:18:08.4 & (x) 0.6 & 359.97359 & 1.249106 & 3.9(H), 5.1(He) & 0.111 & \makecell[tl]{\citet{Skinner1987}$^{y}$, \citet{Tomsick2007}$^{c}$\\ \citet{Galloway2020}$^{d}$} \\
44 & cUCXB & 1990 & SLX 1744-299 & 17:47:25.89 & -30:00:01.6 & (o/i) 0.4 & 359.295259 & -0.889669 & <7.2 (H or He?) & -0.112 & \makecell[tl]{\citet{Skinner1990}$^{y}$, \citet{Zolotukhin2011}$^{c}$\\ \citet{Chelovekov2017}$^{d}$} \\
45 & cUCXB & 1990 & 1RXS J180408.9-342058 & 18:04:08.37 & -34:20:51.19 & (r) 0.006,0.02 & 357.297382 & -6.125046 & <5.8 (He) & -0.619 & \makecell[tl]{\citet{Voges1999}$^{y}$, \citet{Deller2015}$^{c}$\\ \citet{Chenevez2012}$^{d}$} \\
46 & cUCXB & 2003 & IGR J17285-2922 & 17:28:38.86 & -29:21:44.0 & (o) <0.1 & 357.651162 & 2.898254 & assumed 8 & 0.405 & \makecell[tl]{\citet{Walter2004,Barlow2005}$^{y}$, \citet{Torres2010}$^{c}$\\ \citet{Sidoli2011}$^{d}$} \\
47 & cUCXB & 2003 & XMMU J181227.8-181234 & 18:12:27.8 & -18:12:34 & (x) 2 & 12.357731 & 0.033671 & 20.9 (H),  27.2 (He) & 0.016 & \makecell[tl]{\citet{Cackett2006}$^{y}$, \citet{Cackett2006}$^{c}$\\ \citet{Galloway2020}$^{d}$} \\
48 & cUCXB & 1971-1974 & 1M 1716-315 & 17:18:47.02 & -32:10:13.54 & (x) 0.08,0.07 & 354.132917 & 3.069692 & 5.1(H), 6.9 (He) & 0.369 & \makecell[tl]{\citet{Markert1975}$^{y}$, \citet{Jonker2007b}$^{c}$\\ \citet{Jonker2004}$^{d}$} \\
49 & cUCXB & 1976 & 1A 1744-361 & 17:48:13.15 & -36:07:57.02 & (r) 0.014, 0.30 & 354.119972 & -4.192072 & <7 (H), <9.1 (He) & -0.665 & \makecell[tl]{\citet{Davison1976}$^{y}$, \citet{Rupen2003}$^{c}$\\ \citet{Galloway2020}$^{d}$} \\

\end{longtable}\end{ThreePartTable}\end{landscape}
\twocolumn

\onecolumn
\begin{landscape}\begin{ThreePartTable}\begin{TableNotes}\scriptsize\item \textbf{Notes. }In the references column (11), “\textit{p}” refers to \Po, “\textit{F}” to X-ray flux, “\textit{m}” to magnitude, “\textit{N}” to \Nh, and “\textit{E}” to $E(B-V)$. \end{TableNotes}\setlength{\tabcolsep}{1.pt}\setlength{\LTcapwidth}{\textwidth}\scriptsize\centering\begin{longtable}{lllllllllll}\caption{\label{table:table2} Main observational properties.}\\\hline\hline\\[-0.15cm](1) & (2) & (3) &(4) &(5)&(6)&(7)&(8) &(9)&(10)& (11)\\ID & Type & Name & P/T & Accretor & \Po & \Fx~[2--10 keV] & mag & \Nh & GC~$E(B-V)$ & References \\&  &  &  &  & (min) & ($\flux$) &  & ($\times 10^{22}$ \nh) & (mag) &  \\[0.1cm]\endfirsthead\caption{continued.}\\\hline\hline\\[-0.15cm](1) & (2) & (3) &(4) &(5)&(6)&(7)&(8) &(9)&(10)& (11)\\ID & Type & Name & P/T & Accretor & \Po & \Fx~[2--10 keV] & mag & \Nh & GC~$E(B-V)$ & References \\&  &  &  &  & (min) & ($\flux$) &  & ($\times 10^{22}$ \nh) & (mag) &  \\\hline\\[-0.15cm]\endhead\hline\insertTableNotes\endfoot\hline\\\endlastfoot\hline\\[-0.15cm]

1 & UCXB & 4U 1820-303 (NGC 6624) & P & NS & 11.42 & 5.02$\times10^{-09}$ & B=18.7 & 0.16$\pm$0.003 & 0.28 $\pm$ 0.03 & \makecell[tl]{\citet{Stella1987}$^{p}$,\citet{Costantini2012}$^{F}$,\citet{King1993a}$^{m}$\\\citet{Sidoli2001}$^{N}$,\citet{Harris2010}$^{E}$} \\
2 & UCXB & 4U 0513-40 (NGC 1851) & P & NS & 17 & 1.33$\times10^{-10}$ & B=21.1 & 0.026 & 0.02$\pm$0.01 & \makecell[tl]{\citet{Zurek2009}$^{p}$,\citet{Fiocchi2011}$^{F}$,\citet{Deutsch1996}$^{m}$\\\citet{Sidoli2001}$^{N}$,\citet{Harris2010}$^{E}$} \\
3 & UCXB & 2S 0918-549 & P & NS & 17.4 & 1.12$\times10^{-10}$ & V = 20.98$\pm$0.07 & 0.3 &  & \makecell[tl]{\citet{Zhong2011}$^{p}$,\citet{Juett2003b}$^{F}$,\citet{Chevalier1987}$^{m}$\\\citet{Juett2003b}$^{N}$} \\
4 & UCXB & 4U 1543-624 & P & NS & 18.2 & 7.27$\times10^{-10}$ & r' = 20.42$\pm$0.03 & 0.35 &  & \makecell[tl]{\citet{Wang2004}$^{p}$,\citet{Juett2003b}$^{F}$,\citet{Wang2004}$^{m}$\\\citet{Juett2003b}$^{N}$} \\
5 & UCXB & 4U 1850-087 (NGC 6712) & P & NS & 20.6 & 1.04$\times10^{-10}$ & B=21.2$\pm$0.2 & 0.4$\pm$0.2 & 0.45$\pm$0.04 & \makecell[tl]{\citet{Homer1996}$^{p}$,\citet{Juett2005}$^{F}$,\citet{Auriere1992}$^{m}$\\\citet{Sidoli2006}$^{N}$,\citet{Harris2010}$^{E}$} \\
6 & UCXB & M15 X-2 (M15/NGC 7078) & P & NS & 22.5806 & 6.71$\times10^{-11}$ & U=18.6 & <0.034 & 0.1$\pm$0.01 & \makecell[tl]{\citet{Dieball2005}$^{p}$,\citet{White2001}$^{F}$,\citet{DeMarchi1994,White2001}$^{m}$\\\citet{White2001}$^{N}$,\citet{Harris2010}$^{E}$} \\
7 & UCXB & 47 Tuc X-9 (47 Tuc/NGC 104) & P &  & 28.18 & 4.73$\times10^{-12}$ & U=19.4 & 0.013 & 0.04$\pm$0.02 & \makecell[tl]{\citet{Bahramian2017}$^{p}$,\citet{Bahramian2017}$^{F}$,\citet{Paresce1992}$^{m}$\\\citet{Heinke2005}$^{N}$,\citet{Harris2010,Salaris2007}$^{E}$} \\
8 & UCXB & IGR J17062-6143 & P(QP) & NS(AMXP) & 37.9701 & 5.53$\times10^{-11}$ & g'=20.23$\pm$0.16 & 0.26 &  & \makecell[tl]{\citet{Strohmayer2018}$^{p}$,\citet{Keek2017}$^{F}$,\citet{HernandezSantisteban2019}$^{m}$\\\citet{Degenaar2017}$^{N}$} \\
9 & UCXB & XTE J1807-294 & T & NS(AMXP) & 40.0741 & 1.19$\times10^{-10}$ & $^{q}$V>24.3 & 0.46$^{+0.03}_{-0.02}$ &  & \makecell[tl]{\citet{Markwardt2003}$^{p}$,\citet{Campana2003}$^{F}$,\citet{DAvanzo2009}$^{m}$\\\citet{Campana2003}$^{N}$} \\
10 & UCXB & 4U 1626-67 & P & NS & 41.538 & 2.96$\times10^{-10}$ & V=18.54$\pm$0.11 & 0.13$^{+0.04}_{-0.03}$ &  & \makecell[tl]{\citet{Middleditch1981}$^{p}$,\citet{Hemphill2021}$^{F}$,\citet{McClintock1977}$^{m}$\\\citet{Krauss2007}$^{N}$} \\
11 & UCXB & XTE J1751-305 & T & NS(AMXP) & 42.42235 & 1.27$\times10^{-09}$ & R>23.1 & 1.000$^{+0.023}_{-0.011} $&  & \makecell[tl]{\citet{Markwardt2002}$^{p}$,\citet{Gierlinski2005}$^{F}$,\citet{Jonker2003b}$^{m}$\\\citet{Gierlinski2005}$^{N}$} \\
12 & UCXB & XTE J0929-314 & T & NS(AMXP) & 43.57910 & 2.70$\times10^{-10}$ & \makecell[tl]{V=18.63$\pm$0.03\\$^{q}$V=27.58$\pm$0.12} & 0.076 &  & \makecell[tl]{\citet{Galloway2002}$^{p}$,\citet{Juett2003a}$^{F}$,\citet{Giles2005,Monelli2005}$^{m}$\\\citet{Juett2003a}$^{N}$} \\
13 & UCXB & MAXI J0911-655 (NGC 2808) & T & NS(AMXP) & 44.332218 & 1.37$\times10^{-10}$ &  & 0.25$\pm$0.01 & 0.22$\pm$0.02 & \makecell[tl]{\citet{Sanna2017}$^{p}$,\citet{Sanna2017}$^{F}$,\\\citet{Sanna2017}$^{N}$,\citet{Harris2010}$^{E}$} \\
14 & UCXB & IGR J16597-3704 (NGC 6256) & T & NS(AMXP) & 45.97 & 1.97$\times10^{-10}$ &  & 0.82$\pm$0.10 & 1.09$\pm$0.11 & \makecell[tl]{\citet{Sanna2018}$^{p}$,\citet{Sanna2018}$^{F}$,\\\citet{Sanna2018}$^{N}$,\citet{Harris2010}$^{E}$} \\
15 & UCXB & 4U 1916-053 & P & NS & 49.75 & 5.57$\times10^{-10}$ & V=20.99$\pm$0.05, & 0.55$\pm$0.04 &  & \makecell[tl]{\citet{Walter1982,White1982a}$^{p}$,\citet{Iaria2006}$^{F}$,\citet{Grindlay1987}$^{m}$\\\citet{Iaria2021}$^{N}$} \\
16 & UCXB & 4U 0614+091 & P & NS & 51.3 & 1.77$\times10^{-09}$ & V = 18.87$\pm$0.08 & 0.3 &  & \makecell[tl]{\citet{Shahbaz2008}$^{p}$,\citet{Lin2018}$^{F}$,\citet{Machin1990}$^{m}$\\\citet{Madej2010}$^{N}$} \\
17 & UCXB & Swift J1756.9-2508 & T & NS(AMXP) & 54.7017 & 1.54$\times10^{-09}$ &  & 8.14$^{+0.14}_{-0.15}$ &  & \makecell[tl]{\citet{Krimm2007}$^{p}$,\citet{Sanna2018a}$^{F}$,\\\citet{Sanna2018a}$^{N}$} \\
18 & UCXB & NGC 6440 X-2 (NGC 6440) & T & NS(AMXP) & 57.3 & 1.58$\times10^{-10}$ & \makecell[tl]{g'>22\\$^{q}$V>21.0} & 0.69$\pm$0.06 & 1.07$\pm$0.10 & \makecell[tl]{\citet{Altamirano2010}$^{p}$,\citet{Heinke2010}$^{F}$,\citet{Heinke2010}$^{m}$\\\citet{Heinke2010}$^{N}$,\citet{Harris1996}$^{E}$} \\
19 & UCXB & MAXI J1957+032 & T & NS(AMXP
) & 60.8841 & 4.91$\times10^{-10}$ & \makecell[tl]{R=18.27$\pm$0.06\\$^{q}$R=21.4$\pm$0.2} & 0.17 &  & \makecell[tl]{\citet{Sanna2022}$^{p}$,\citet{MataSanchez2017a}$^{F}$,\citet{Guver2015,MataSanchez2017a}$^{m}$\\\citet{MataSanchez2017a}$^{N}$} \\
20 & UCXB & IGR J17494-3030 & T & NS(AMXP) & 74.9445 & 9.35$\times10^{-11}$ &  & 1.87 &  & \makecell[tl]{\citet{Ng2021}$^{p}$,\citet{ArmasPadilla2013c}$^{F}$,\\\citet{ArmasPadilla2013c}$^{N}$} \\
21 & spLMXB & HETE J1900.1–2455 & T(LO) & NS(AMXP
) & 83.2543 & 3.92$\times10^{-10}$ & V=18.09$\pm$0.03 & 0.105 &  & \makecell[tl]{\citet{Kaaret2006}$^{p}$,\citet{Papitto2013}$^{F}$,\citet{Steeghs2005b}$^{m}$\\\citet{Papitto2013}$^{N}$} \\
22 & spLMXB & 1E 1603.6+2600 & P & NS & 111.04 & 4.19$\times10^{-12}$ & R=19.4 & 0.15$\pm$0.03 &  & \makecell[tl]{\citet{Morris1990}$^{p}$,\citet{Jonker2003a}$^{F}$,\citet{Morris1990}$^{m}$\\\citet{Jonker2003a}$^{N}$} \\
23 & spLMXB & IGR J17379–3747 & T & NS(AMXP) & 111.65 & 1.23$\times10^{-10}$ & R=21.5$\pm$0.3 & 0.90$\pm$0.03 &  & \makecell[tl]{\citet{Sanna2018b,Strohmayer2018a}$^{p}$,\citet{VandenEijnden2018b}$^{F}$,\citet{Curran2011}$^{m}$\\\citet{Sanna2018b}$^{N}$} \\
24 & spLMXB & 4U 1812-12 & P & NS & 114 & 3.80$\times10^{-10}$ & r=22.7 & 1.50 &  & \makecell[tl]{\citet{ArmasPadilla2022}$^{p}$,\citet{Muno2005b}$^{F}$,\citet{ArmasPadilla2020}$^{m}$\\\citet{Tarana2006,Barret2003}$^{N}$} \\
25 & cUCXB & 4U 1728-34 & P & NS &  & 2.08$\times10^{-09}$ & K=15.1 & 4.5$\pm$0.1 &  & \makecell[tl]{\citet{Wang2019}$^{F}$,\citet{Marti1998}$^{m}$\\\citet{Wang2019}$^{N}$} \\
26 & cUCXB & NGC 6652 B (NGC 6652) & T & NS &  & 7.52$\times10^{-13}$ & B = 20.4 & 0.05 & 0.09$\pm$0.01 & \makecell[tl]{\citet{Stacey2012}$^{F}$,\citet{Deutsch1998}$^{m}$\\\citet{Coomber2011}$^{N}$,\citet{Harris2010}$^{E}$} \\
27 & cUCXB & 4U 1822-00 & P & NS &  & 5.80$\times10^{-10}$ & r'=21.58$\pm$0.08 & 0.97$\pm$0.18 &  & \makecell[tl]{\citet{Hertz1984}$^{F}$,\citet{Wang2004}$^{m}$\\\citet{Juett2005}$^{N}$} \\
28 & cUCXB & 4U 1905+000 & T(LO) & NS &  & 1.57$\times10^{-10}$ & V = 20.5$\pm$0.04 & 0.3$\pm$0.1 &  & \makecell[tl]{\citet{Christian1997}$^{F}$,\citet{Chevalier1985b}$^{m}$\\\citet{Chevalier1990}$^{N}$} \\
29 & cUCXB & RX J1709.5-2639 (NGC 6293) & T & NS &  & 9.10$\times10^{-10}$ & \makecell[tl]{R=20.5$\pm$0.1\\$^{q}$R=22.24$\pm$0.03} & 0.29$\pm$0.01 & 0.36$\pm$0.04 & \makecell[tl]{\citet{Ludlam2017}$^{F}$,\citet{Jonker2004a}$^{m}$\\\citet{Degenaar2013}$^{N}$,\citet{Harris2010}$^{E}$} \\
30 & cUCXB & AX J1754.2-2754 & P(QP) & NS &  & 9.55$\times10^{-12}$ & Ks=18.12$\pm$0.15 & 2.93$\pm$0.06 &  & \makecell[tl]{\citet{ArmasPadilla2013b}$^{F}$,\citet{Shaw2017}$^{m}$\\\citet{ArmasPadilla2013b}$^{N}$} \\
31 & cUCXB & 1RXS J171824.2-402934 & P & NS &  & 3.75$\times10^{-12}$ &  & 1.9$\pm$0.1 &  & \makecell[tl]{\citet{ArmasPadilla2013b}$^{F}$,\\\citet{ArmasPadilla2013b}$^{N}$} \\
32 & cUCXB & 1RXH J173523.7-354013 & P & NS &  & 2.30$\times10^{-12}$ & V=21.2 & 1.4$\pm$0.1 &  & \makecell[tl]{\citet{ArmasPadilla2013b}$^{F}$,\citet{Degenaar2010b}$^{m}$\\\citet{ArmasPadilla2013b}$^{N}$} \\
33 & cUCXB & XMMU J174716.1-281048 & T(LO) & NS &  & 6.40$\times10^{-12}$ & H=15.3$\pm$0.1 & 8.9$\pm$0.5 &  & \makecell[tl]{\citet{degenaar2011_burstxmmsource}$^{F}$,\citet{Degenaar2007}$^{m}$\\\citet{DelSanto2007}$^{N}$} \\
34 & cUCXB & SAX J1806.5-2215 & T
(LO) & NS &  & 6.60$\times10^{-10}$ & \makecell[tl]{K=17.25$\pm$0.03\\$^{q}$K>18.2} & 4.2$\pm$0.6 &  & \makecell[tl]{\citet{delsanto2012}$^{F}$,\citet{Kaur2017}$^{m}$\\\citet{delsanto2012}$^{N}$} \\
35 & cUCXB & AX J1538.3-5541 & P(QP) &  &  & 2.57$\times10^{-11}$ & K=18 & 7.7$\pm$0.5 &  & \makecell[tl]{\citet{Degenaar2012d}$^{F}$,\citet{Anderson2014}$^{m}$\\\citet{Degenaar2012d}$^{N}$} \\
36 & cUCXB & Swift J0840.7-3516 & T &  &  & 9.91$\times10^{-10}$ & \makecell[tl]{R=16.3$\pm$0.1\\$^{q}$r=21} & 0.46$^{+0.13}_{-0.10}$ &  & \makecell[tl]{\citet{CotiZelati2021}$^{F}$,\citet{Melandri2020,CotiZelati2021}$^{m}$\\\citet{CotiZelati2021}$^{N}$} \\
37 & cUCXB & 1A 1246-588 & P & NS &  & 9.10$\times10^{-11}$ & V=19.45 & 0.5 &  & \makecell[tl]{\citet{IntZand2008}$^{F}$,\citet{Bassa2006}$^{m}$\\\citet{IntZand2008}$^{N}$} \\
38 & cUCXB & SAX J1712.6-3739 & P & NS &  & 1.29$\times10^{-10}$ & V=23.93$\pm$0.26 & 1.3 &  & \makecell[tl]{\citet{Lin2020}$^{F}$,\citet{Wiersema2009}$^{m}$\\\citet{Fiocchi2008b}$^{N}$} \\
39 & cUCXB & 1RXS J170854.4-321857 & P(QP) & NS &  & 4.57$\times10^{-11}$ & K > 20.1 & 0.34$\pm$0.04 &  & \makecell[tl]{\citet{ArmasPadilla2019b}$^{F}$,\citet{Revnivtsev2013}$^{m}$\\\citet{ArmasPadilla2019b}$^{N}$} \\
40 & cUCXB & 4U 1722-30 (Terzan 2) & P & NS &  & 5.74$\times10^{-10}$ &  & 1.1 & 1.87$\pm$0.19 & \makecell[tl]{\citet{Lin2018}$^{F}$,\\\citet{Kashyap2022}$^{N}$,\citet{Harris2010,Valenti2010}$^{E}$} \\
41 & cUCXB & IGR J17254-3257 & P & NS &  & 2.77$\times10^{-11}$ &  & 1.79 &  & \makecell[tl]{\citet{Chenevez2007}$^{F}$,\\\citet{Chenevez2007}$^{N}$} \\
42 & cUCXB & SLX 1735-269 & P & NS &  & 3.45$\times10^{-10}$ & R=21.31$\pm$0.12 & 1.70$\pm$0.05 &  & \makecell[tl]{\citet{Wilson2003}$^{F}$,\citet{Zolotukhin2011}$^{m}$\\\citet{Wilson2003}$^{N}$} \\
43 & cUCXB & SLX 1737-282 & P & NS &  & 5.53$\times10^{-11}$ &  & 1.9+/-0.1 &  & \makecell[tl]{\citet{ArmasPadilla2018}$^{F}$,\\\citet{ArmasPadilla2018,IntZand2002}$^{N}$} \\
44 & cUCXB & SLX 1744-299 & P & NS &  & 2.08$\times10^{-10}$ & I=23.37$\pm$0.28 & 3.5 &  & \makecell[tl]{\citet{Mori2005}$^{F}$,\citet{Zolotukhin2011}$^{m}$\\\citet{Mori2005}$^{N}$} \\
45 & cUCXB & 1RXS J180408.9-342058 & T & NS &  & 1.97$\times10^{-09}$ & V=17.71$\pm$0.02 & 0.48+/-0.02 &  & \makecell[tl]{\citet{Degenaar2016}$^{F}$,\citet{Baglio2016}$^{m}$\\\citet{Degenaar2016}$^{N}$} \\
46 & cUCXB & IGR J17285-2922 & T &  &  & 1.04$\times10^{-10}$ & I=18.24 & 0.99+/-0.05 &  & \makecell[tl]{\citet{Stoop2021}$^{F}$,\citet{Stoop2021}$^{m}$\\\citet{Stoop2021}$^{N}$} \\
47 & cUCXB & XMMU J181227.8-181234 & T & NS &  & 2.85$\times10^{-09}$ &  & 12.8$\pm$0.3 &  & \makecell[tl]{\citet{Cackett2006}$^{F}$,\\\citet{Cackett2006}$^{N}$} \\
48 & cUCXB & 1M 1716-315 & T(LO) & NS &  & 1.07$\times10^{-09}$ & I>23.5 & 0.21$\pm$0.06 &  & \makecell[tl]{\citet{Jonker2007b}$^{F}$,\citet{Jonker2007b}$^{m}$\\\citet{Christian1997}$^{N}$} \\
49 & cUCXB & 1A 1744-361 & T & NS &  & 1.4$\times10^{-9}$ &  & 0.41$\pm$0.01 &  & \makecell[tl]{\citet{Gavriil2012}$^{F}$,\\\citet{Gavriil2012}$^{N}$} \\

\end{longtable}\end{ThreePartTable}\end{landscape}
\twocolumn

\onecolumn
\begin{landscape}\begin{ThreePartTable}\begin{TableNotes}\scriptsize\item \textbf{Notes. }We indicate whether the features are in absorption ($-$) or in emission ($+$) whenever it could be established from the original study. In the references column (12), “\textit{B}” refers to short burst, “\textit{IB}” to intermediate long burst, “\textit{SB}” to super bursts, “\textit{X}” to X-ray information,“\textit{o}” to optical information,“\textit{UV}” to UV information, “\textit{o/x}” to low optical–to–X-ray luminosity ratio ($L_{\rm o}/$\lx$<<$), and “\textit{Lp}” to low persistent X-ray luminosities (\lx$^\mathrm{Pers}<<$) \end{TableNotes}\setlength{\tabcolsep}{1.pt}\setlength{\LTcapwidth}{\textwidth}\scriptsize\centering\begin{longtable}{llllllllllll}\caption{\label{table:table3} Additional multiwavelength properties.}\\\hline\hline\\[-0.15cm](1) & (2) & (3) &(4) &(5)&(6)&(7)&(8) &(9)&(10)& (11)&(12)\\ID & Type & Name & short-B & IB & SB & X-ray & Optical & UV & $L_{\rm o}/$\lx$<<$ & \lx$^\mathrm{Pers}<<$ &References\\\\[-0.15cm]\endfirsthead\caption{continued.}\\\hline\hline\\[-0.15cm](1) & (2) & (3) &(4) &(5)&(6)&(7)&(8) &(9)&(10)& (11)&(12)\\ID & Type & Name & short-B & IB & SB & X-ray & Optical & UV & $L_{\rm o}/$\lx$<<$ & \lx$^\mathrm{Pers}<<$ &References\\\hline\\[-0.15cm]\endhead\hline\insertTableNotes\endfoot\hline\\\endlastfoot\hline\\[-0.15cm]

1 & UCXB & 4U 1820-303 (NGC 6624) & \checkmark &  & \checkmark & $-$O~\textsc{vii}, $-$O~\textsc{viii}, $-$Ne~\textsc{ix} &  &  &  &  & \makecell[tl]{\citet{Clark1977}$^{B}$, \citet{Strohmayer2002a}$^{SB}$\\\citet{Futamoto2004,Cackett2008}$^{X}$} \\
2 & UCXB & 4U 0513-40 (NGC 1851) & \checkmark &  &  & $-$O &  &  & \checkmark &  & \makecell[tl]{\citet{Forman1976b,Galloway2020}$^{B}$\\\citet{Juett2005}$^{X}$, \citet{Deutsch2000}$^{o/x}$} \\
3 & UCXB & 2S 0918-549 & \checkmark & \checkmark &  & Ne & Featureless &  & \checkmark & \checkmark & \makecell[tl]{\citet{Jonker2001b}$^{B}$,\citet{IntZand2005}$^{IB}$,\citet{Juett2001}$^{X}$\\\citet{Nelemans2004,Nelemans2006}$^{o}$,\citet{Juett2001}$^{o/x}$\\\citet{Jonker2001b}$^{Lp}$} \\
4 & UCXB & 4U 1543-624 & \checkmark &  &  & Ne/O & $+$C, $+$O &  & \checkmark & \checkmark & \makecell[tl]{\citet{Serino2018}$^{B}$,\citet{Juett2003b}$^{X}$,\citet{Nelemans2004}$^{o}$\\\citet{Juett2001}$^{o/x}$,\citet{Wang2004}$^{Lp}$} \\
5 & UCXB & 4U 1850-087 (NGC 6712) & \checkmark & \checkmark &  & Ne & Featureless & Featureless & \checkmark & \checkmark & \makecell[tl]{\citet{Swank1976,Hoffman1980}$^{B}$\citet{IntZand2014b}$^{IB}$\\\citet{Juett2001}$^{X}$\\\citet{Auriere1992,Downes1996}$^{o,uv}$\\\citet{Juett2001}$^{o/x}$\citet{IntZand2007}$^{Lp}$} \\
6 & UCXB & M15 X-2 (M15/NGC 7078) & \checkmark & \checkmark &  &  &  & $+$C~\textsc{iv}?,$+$He~\textsc{ii}? &  & \checkmark & \makecell[tl]{\citet{vanParadijs1990b,Smale2001}$^{B}$\citet{Galloway2020}$^{IB}$\\\citet{Dieball2005}$^{uv}$,\citet{IntZand2007}$^{Lp}$} \\
7 & UCXB & 47 Tuc X-9 (47 Tuc/NGC 104) &  &  &  & \makecell[tl]{$+$O~\textsc{vii}, $+$O~\textsc{viii}\\Bump at $\sim$0.3~keV} & Featureless & $+$C~\textsc{iv} &  & \checkmark & \makecell[tl]{\citet{Heinke2005, Bahramian2017}$^{X}$\\\citet{Tudor2018}$^{o}$\citet{Knigge2008}$^{uv}$\\\citet{Bahramian2017, Bahramian2022}$^{Lp}$} \\
8 & UCXB & IGR J17062-6143 &  & \checkmark & \checkmark & \makecell[tl]{$+$O\\Absorption at $\sim$3.4~keV (Ca or Ti)} & Featureless &  &  & \checkmark & \makecell[tl]{\citet{Degenaar2012a}$^{IB}$, \citet{Negoro2015}$^{SB}$\\\citet{VandenEijnden2018, Bult2021}$^{X}$\\\citet{HernandezSantisteban2019}$^{o}$,\citet{Degenaar2017}$^{Lp}$} \\
9 & UCXB & XTE J1807-294 &  &  &  &  &  &  &  &  &  \\
10 & UCXB & 4U 1626-67 &  &  &  & \makecell[tl]{$+$Ne~\textsc{x}, $+$Ne~\textsc{ix}, $+$O~\textsc{viii},\\$+$O~\textsc{vii}, $+$Fe$-$L} & $+$C, $+$O & \makecell[tl]{Ly$\alpha$, C~\textsc{iii}, O~\textsc{iv}, O~\textsc{v},\\Si~\textsc{iv}, C~\textsc{iv}, O~\textsc{iii}} &  &  & \makecell[tl]{\citet{Angelini1995,Schulz2001,Krauss2007}$^{X}$\\\citet{Werner2006,Nelemans2006}$^{o}$\citet{Homer2002}$^{uv}$} \\
11 & UCXB & XTE J1751-305 &  &  &  &  &  &  &  &  & \\
12 & UCXB & XTE J0929-314 &  &  &  & $-$O & \makecell[tl]{Emission at $\sim$4640$\lambda$\\(C or N)} &  &  &  & \makecell[tl]{\citet{Juett2003a}$^{X}$,\citet{Nelemans2006}$^{o}$} \\
13 & UCXB & MAXI J0911-655 (NGC 2808) &  & \checkmark &  &  &  &  &  &  & \makecell[tl]{\citet{Nakajima2020}$^{IB}$} \\
14 & UCXB & IGR J16597-3704 (NGC 6256) &  &  &  &  &  &  &  &  &  \\
15 & UCXB & 4U 1916-053 & \checkmark &  &  & \makecell[tl]{$-$Ne~\textsc{x}, $-$Mg~\textsc{xii}, $-$Si~\textsc{xiv},\\$-$S~\textsc{xvi}, $-$O~\textsc{viii}, $-$Ca~\textsc{xx}} & He~\textsc{i}, He~\textsc{ii}, N~\textsc{ii}, N~\textsc{iii} &  &  & \checkmark & \makecell[tl]{\citet{Becker1977}$^{B}$\\\citet{Iaria2006,Gambino2019,Iaria2021}$^{X}$\\\citet{Nelemans2006}$^{o}$,\citet{IntZand2007}$^{Lp}$} \\
16 & UCXB & 4U 0614+091 & \checkmark & \checkmark & \checkmark & Ne, O & $+$C, $+$O, $+$N & Featureless & \checkmark & \checkmark & \makecell[tl]{\citet{Lewin1976}$^{B}$\citet{Kuulkers2010}$^{IB}$\\\citet{Kuulkers2005}$^{SB}$\citet{Juett2001, Madej2010}$^{X}$\\\citet{Machin1990,Nelemans2004,Madej2013}$^{o}$\\\citet{Madej2013}$^{uv}$\citet{Juett2001}$^{o/x}$\citet{IntZand2007}$^{Lp}$} \\
17 & UCXB & Swift J1756.9-2508 &  &  &  &  &  &  &  &  &  \\
18 & UCXB & NGC 6440 X-2 (NGC 6440) &  &  &  &  &  &  &  &  &  \\
19 & UCXB & MAXI J1957+032 &  &  &  &  & Featureless &  &  &  & \makecell[tl]{\citet{MataSanchez2017a}$^{o}$} \\
20 & UCXB & IGR J17494-3030 &  &  &  &  &  &  &  &  &  \\
21 & spLMXB & HETE J1900.1–2455 & \checkmark &  &  & \makecell[tl]{Excess at $\sim$1~keV\\(Fe-L$\alpha$ and/or Ne-K$\alpha$)} & \makecell[tl]{$+$H$\alpha$, $+$He~\textsc{ii}\\Bowen blend of N~\textsc{iii} and C~\textsc{iii}} &  &  &  & \makecell[tl]{\citet{Vanderspek2005,Galloway2020}$^{B}$\\\citet{Papitto2013}$^{X}$,\citet{Elebert2007,Steeghs2005b}$^{o}$} \\
22 & spLMXB & 1E 1603.6+2600 & \checkmark &  &  & \makecell[tl]{Emission at $\sim$0.5~keV and $\sim$0.65~keV\\(N~\textsc{vii} and/or O~\textsc{viii})} & \makecell[tl]{$+$H$\alpha$, $+$H$\beta$, $+$He~\textsc{ii}\\$-$He~\textsc{i}, $-$Ca~\textsc{ii}} &  &  & \checkmark & \makecell[tl]{\citet{Mukai2001}$^{B}$,\citet{Hakala2005}$^{B,X}$\\\citet{Morris1990}$^{o}$,\citet{IntZand2007}$^{Lp}$} \\
23 & spLMXB & IGR J17379–3747 & \checkmark &  &  &  &  &  &  &  & \makecell[tl]{\citet{Chelovekov2006,Chelovekov2010}$^{B}$} \\
24 & spLMXB & 4U 1812-12 & \checkmark &  &  &  & Featureless &  & \checkmark & \checkmark & \makecell[tl]{\citet{Murakami1983,Galloway2020}$^{B}$\\\citet{ArmasPadilla2022}$^{o}$,\citet{Bassa2006}$^{o/x}$\\\citet{IntZand2007}$^{Lp}$} \\
25 & cUCXB & 4U 1728-34 & \checkmark &  &  & \makecell[tl]{$+$Mg XI, emission at $\sim$0.5~keV\\Absoprtion at $\sim$0.7~keV} &  &  &  &  & \makecell[tl]{\citet{Hoffman1976,Galloway2010}$^{B}$\\\citet{diSalvo2000,Koliopanos2021}$^{X}$} \\
26 & cUCXB & NGC 6652 B (NGC 6652) & \checkmark &  &  &  & \makecell[tl]{$+$H, $+$He\\$-$Fe, $-$Mg, $-$Na, $-$O} &  &  &  & \makecell[tl]{\citet{Intzand1998}$^{B}$,\citet{Paduano2021}$^{o}$} \\
27 & cUCXB & 4U 1822-00 & \checkmark &  &  &  & Featureless &  & \checkmark &  & \makecell[tl]{\citet{Asai2016}$^{B}$,\citet{Nelemans2006}$^{o}$\\\citet{Juett2005}$^{o/x}$} \\
28 & cUCXB & 4U 1905+000 & \checkmark &  &  &  &  &  & \checkmark &  & \makecell[tl]{\citet{Lewin1976b}$^{B}$,\citet{Chevalier1990}$^{o/x}$} \\
29 & cUCXB & RX J1709.5-2639 (NGC 6293) & \checkmark &  &  & \makecell[tl]{Emission at $\sim$0.6~keV\\(O and/or Fe, enhanced Ne/O)} &  &  &  &  & \makecell[tl]{\citet{Cocchi1998}$^{B}$,\citet{Jonker2003}$^{X}$} \\
30 & cUCXB & AX J1754.2-2754 & \checkmark & \checkmark &  &  &  &  & \checkmark & \checkmark & \makecell[tl]{\citet{Chelovekov2007}$^{B}$\citet{Chenevez2017}$^{IB}$\\\citet{Bassa2008}$^{o/x}$\citet{Bassa2008}$^{Lp}$} \\
31 & cUCXB & 1RXS J171824.2-402934 & \checkmark &  &  &  &  &  &  & \checkmark & \makecell[tl]{\citet{Kaptein2000}$^{B}$,\citet{IntZand2005}$^{Lp}$} \\
32 & cUCXB & 1RXH J173523.7-354013 &  & \checkmark &  &  & $+$H$\alpha$, $+$Ca~\textsc{ii}, $+$O~\textsc{i} &  &  & \checkmark & \makecell[tl]{\citet{Degenaar2010b}$^{IB,o,Lp}$,\citet{ArmasPadilla2013b}$^{Lp}$} \\
33 & cUCXB & XMMU J174716.1-281048 & \checkmark & \checkmark &  &  &  &  &  &  & \makecell[tl]{\citet{Brandt2006}$^{B}$\citet{degenaar2011_burstxmmsource}$^{IB}$} \\
34 & cUCXB & SAX J1806.5-2215 & \checkmark & \checkmark &  &  &  &  &  & \checkmark & \makecell[tl]{\citet{Cornelisse2002a}$^{B,Lp}$,\citet{IntZand2019}$^{IB}$} \\
35 & cUCXB & AX J1538.3-5541 &  &  &  &  &  &  &  &  &  \\
36 & cUCXB & Swift J0840.7-3516 &  &  &  & \makecell[tl]{Absorption at $\sim$3.4~keV\\(Fe K?)} & \makecell[tl]{Blend of C  and O in emission,\\$+$He~\textsc{i}, $+$C~\textsc{ii}} &  &  &  & \makecell[tl]{\citet{Shidatsu2021}$^{X}$,\citet{CotiZelati2021}$^{o}$} \\
37 & cUCXB & 1A 1246-588 & \checkmark & \checkmark &  & $-$O, $-$Ne, $-$Fe & Featureless &  & \checkmark & \checkmark & \makecell[tl]{\citet{Boller1997,Piro1997}$^{B}$,\citet{IntZand2008}$^{IB}$\\\citet{IntZand2008}$^{X}$,\citet{IntZand2008}$^{o}$\\\citet{Bassa2006}$^{o/x}$,\citet{IntZand2007}$^{Lp}$} \\
38 & cUCXB & SAX J1712.6-3739 & \checkmark & \checkmark &  &  &  &  &  & \checkmark & \makecell[tl]{\citet{Cocchi1999}$^{B}$,\citet{Kuulkers2009}$^{IB}$,\citet{IntZand2007}$^{Lp}$} \\
39 & cUCXB & 1RXS J170854.4-321857 &  & \checkmark &  & \makecell[tl]{Overabundances of Ne and Fe\\at $\sim$1~keV} &  &  &  & \checkmark & \makecell[tl]{\citet{IntZand2004,IntZand2005}$^{IB}$\\\citet{ArmasPadilla2019b}$^{X}$,\citet{IntZand2007}$^{Lp}$} \\
40 & cUCXB & 4U 1722-30 (Terzan 2) & \checkmark &  &  &  &  &  &  & \checkmark & \makecell[tl]{\citet{Swank1977,Kuulkers2003}$^{B}$\\\citet{IntZand2007}$^{Lp}$} \\
41 & cUCXB & IGR J17254-3257 & \checkmark & \checkmark &  &  &  &  &  & \checkmark & \makecell[tl]{\citet{Brandt2006b}$^{B}$\citet{Chenevez2007}$^{IB}$\\\citet{IntZand2007}$^{Lp}$} \\
42 & cUCXB & SLX 1735-269 & \checkmark & \checkmark & \checkmark &  &  &  &  & \checkmark & \makecell[tl]{\citet{Bazzano1997}$^{B}$\citet{Molkov2005}$^{IB}$\\\citet{Suzuki2005}$^{SB}$,\citet{IntZand2007}$^{Lp}$} \\
43 & cUCXB & SLX 1737-282 &  & \checkmark &  &  &  &  &  & \checkmark & \makecell[tl]{\citet{IntZand2002,Falanga2008}$^{IB}$\\\citet{IntZand2007}$^{Lp}$} \\
44 & cUCXB & SLX 1744-299 &  & \checkmark &  &  &  &  &  & \checkmark & \makecell[tl]{\citet{Pavlinsky1994,IntZand2007,Alizai2020}$^{IB}$\\\citet{IntZand2007}$^{Lp}$} \\
45 & cUCXB & 1RXS J180408.9-342058 & \checkmark &  &  & $+$N VII, $+$O VII, $+$ O VIII & $+$He~\textsc{ii} &  &  &  & \makecell[tl]{\citet{Chenevez2012,Wijnands2017}$^{B}$\\\citet{Ludlam2016,Degenaar2016}$^{X}$,\citet{Baglio2016}$^{o}$} \\
46 & cUCXB & IGR J17285-2922 &  &  &  & Negative residuals at $\leq$1~keV & Featureless &  &  &  & \makecell[tl]{\citet{Sidoli2011}$^{X}$,\citet{Stoop2021}$^{o}$} \\
47 & cUCXB & XMMU J181227.8-181234 & \checkmark &  &  &  &  &  &  &  & \makecell[tl]{\citet{Goodwin2019}$^{B}$} \\
48 & cUCXB & 1M 1716-315 & \checkmark & \checkmark &  &  &  &  &  &  & \makecell[tl]{\citet{Hoffman1978}$^{B}$,\citet{Markert1976,Tawara1984}$^{IB}$} \\
49 & cUCXB & 1A 1744-361 & \checkmark &  &  &  &  &  &  &  & \makecell[tl]{\citet{Emelyanov2001,Bhattacharyya2006}$^{B}$} \\

\end{longtable}\end{ThreePartTable}\end{landscape}
\twocolumn

\end{appendix}

\end{document}